\documentclass[%
 reprint,
 amsmath,amssymb,
 aps,
]{revtex4-2}

\usepackage{graphicx}
\usepackage{dcolumn}
\usepackage{bm}
\usepackage{hyperref}

\begin{document}

\preprint{APS/123-QED}

\title{Phase transition in large language models and the criticality of natural languages}

\author{Kai Nakaishi}
\affiliation{%
 Center for Advanced Intelligence Project, RIKEN
}
\affiliation{%
 National Institute for Japanese Language and Linguistics
}
\author{Yoshihiko Nishikawa}%
\affiliation{%
 Department of Physics, Nagoya University
}%
\author{Koji Hukushima}%
\affiliation{%
 Department of Multidisciplinary Sciences, The University of Tokyo
}%
\affiliation{%
 Komaba Institute for Science, The University of Tokyo
}%

\date{\today}

\begin{abstract}
Generation of text and speech in natural languages can be modeled as a stochastic process. This idea dates back to the seminal work of Markov and, later, to that of Shannon and also underlies the recent development of large language models (LLMs). The stochastic processes corresponding to natural languages should be distinct from those that generate nonlinguistic sequences. One of the features that discriminate linguistic and nonlinguistic sequences is power-law behavior, which is universally observed across different languages. In statistical physics, such behavior suggests that natural languages are \emph{critical}: They lie near a phase transition point in a parametrized space of stochastic processes. However, testing this conjecture is not straightforward. A phase transition, even if it exists, cannot be directly observed in real-world natural languages because they do not have any controllable parameters. Here, we use LLMs as controllable effective models of natural languages. Through statistical analyses of texts generated by LLMs, we find that, when a parameter analogous to physical temperature is varied, LLMs undergo a phase transition. The transition separates a low-temperature phase with complex repetitive structures in generated texts from a high-temperature phase in which LLMs generate incomprehensible texts. At the critical point between these phases, generated texts display the power-law behavior similar to that of natural languages and most closely resemble natural languages as measured by a standard metric in natural language processing. These findings strongly suggest that natural languages are indeed critical.
\end{abstract}

\maketitle


Text and speech can be naturally viewed as a symbol sequence generated by a stochastic process in which the next word following a given sequence is drawn from a probability distribution that depends on syntactic, semantic, discourse, and other factors. For example, given a context \emph{Alice was beginning to}, the probability assigned to \emph{get} as the next word is much higher than that for \emph{gets}, which is grammatically incorrect. This idea of regarding natural languages as stochastic processes dates back to the seminal work of Markov~\cite{Markov_2006}, which showed that characters in a text are not independent, and that of Shannon~\cite{shannon1951prediction}, which quantified the information carried by a character. More recently, this idea underpins the development of language models, which are trained to predict the next token.

We take this idea one step further and ask what the nature of natural languages is: What features characterize stochastic processes corresponding to natural languages among all possible processes generating symbol sequences? Where do such processes lie in the space of possible stochastic processes?

A distinctive statistical feature of natural languages is \emph{power-law behavior}. The most well-known example is Zipf's law~\cite{zipf1935psychobiology}, which states that word frequencies follow a power law as a function of their ranks. Heaps' law~\cite{herden1964quantitative, heaps1978information} describes a power-law relation between the number of distinct words and the length of a document. The decay of correlations between linguistic elements such as phones, characters, and words also follows a power law with the distance between elements~\cite{Li1989mutual, Ebeling1994-om, Ebeling1995-om, Tanaka-Ishii2016-oj, Lin2017-gs, takahashi2017neural, shen2019mutual, takahashi2019evaluating, Sainburg2019-tn, mikhaylovskiy2023autocorrelations}. Remarkably, these phenomena have been universally observed across corpora, languages, and methodological details. 

A variety of theories have been proposed to account for these phenomena, ranging from simple random processes~\cite{miller1957some} and preferential sampling models~\cite{simon1955class}, to the principle of least effort~\cite{zipf1949human, mandelbrot1965information} and explanations based on underlying hierarchical structure~\cite{Lin2017-gs}. However, it remains unclear whether these mechanisms adequately explain the power-law behavior in natural languages~\cite{piantadosi2014zipf, nakaishi2025rethinking}. 

Alternatively, in statistical physics, power-law behavior implies phase transitions and criticality~\cite{stanley1971introduction}. A phase transition refers to a sharp, qualitative change in the macroscopic behavior of a system that occurs as a control parameter, such as temperature or an external field, is varied. For example, as the temperature increases, a magnet consisting of many magnetic spins undergoes a transition from an ordered phase, in which the spins are aligned, to a disordered phase, in which their orientations are nearly independent. A system is said to be \emph{critical} when it is near a critical point, that is, the point separating the phases. When systems are critical, power-law behavior emerges in various statistical quantities; this phenomenon is known as a critical phenomenon. While the concept of critical phenomena originates in condensed matter physics, it has since been applied to a wide range of systems, including biological systems~\cite{munoz2018colloquium}, natural disasters~\cite{sachs2012black}, economic and financial systems~\cite{bouchaud2024self}, and computation~\cite{langton1990computation}. The power-law behavior in natural languages therefore motivates the conjecture that natural languages are critical: The stochastic process corresponding to a natural language lies near a critical point between two distinct phases in a parametrized space of stochastic processes.

Testing this conjecture is not straightforward. Because natural languages lack experimentally controllable parameters, we cannot directly observe whether a phase transition occurs as a parameter is varied. For this reason, previous attempts to associate natural languages with phase transitions have relied on mathematical models~\cite{ferrer-i-cancho2003least, ferrer-i-cancho2005zipf, ferrer-i-cancho2007global, prokopenko2010phase}. However, these models are oversimplified, and it remains unclear to what extent they provide insight into real-world natural languages.

Large language models (LLMs) can serve as more realistic models of natural languages. An LLM trained on large-scale real-world corpora for a natural language can generate extremely diverse and linguistically natural texts in that language. In addition, unlike natural languages, LLMs have controllable parameters, enabling us to empirically investigate the statistical properties of generated texts as functions of these parameters. In addition, LLMs exhibit a notable behavior as a \emph{temperature} parameter $T$, analogous to physical temperature, is varied. Empirically, texts generated at low temperatures contain repeated patterns~\cite{Holtzman2020The}, such as \emph{The number of schools in the United States\textbackslash{}n\textbackslash{}nThe number of schools in the United States\textbackslash{}n\textbackslash{}n ...}, whereas at high temperatures, texts become incomprehensible, as in \emph{detached speeches tailor hello cellular networks early Symbol dissentuggest table price laced efma ...}. This behavior is reminiscent of a phase transition from an ordered to a disordered phase and suggests a connection between natural languages and phase transitions.

Nevertheless, only a limited number of studies have related this temperature-induced change to phase transitions. Bahamondes~\cite{Bahamondes2023study} observed correlations in LLM-generated texts and argued that a phase transition occurs, yet the experiment was conducted at a limited scale. We also note that the phenomenon discussed in that study would be distinct from those in other studies, including ours, because the transition temperature reported by Bahamondes~\cite{Bahamondes2023study} is significantly lower. Arnold et al.~\cite{arnold2024phase} and Sun and Haghighat~\cite{sun2025phase} introduced quantities analogous to energy and observed singular behavior in these quantities as a function of temperature, indicating the presence of a phase transition. Mikhaylovskiy~\cite{mikhaylovskiy-2025-zipfs} reported that Zipf’s and Heaps’ laws emerge near the transition temperature. These studies, however, have not fully characterized the nature of the phases separated by the transition, nor have they clearly discussed how the phase transitions relate to the criticality of natural languages.

In this study, we perform statistical analyses of texts generated by LLMs and show that the behavior of correlations changes qualitatively at $T \approx 1$, establishing the presence of a phase transition and critical phenomena. We also show, through power-spectrum analyses, that repetitive structures in the low-temperature phase are complex, characterized by a divergent number of distinct periods. We further conduct similar analyses for partially trained models, which have not been examined in the previous studies. The results show that the critical behavior and the repetitive structures emerge as the model learns a natural language, indicating that both are intrinsic to natural languages. Finally, using perplexity, a standard metric in natural language processing, we show that the model most closely matches a natural language dataset near the critical point at the final stage of training.

\begin{figure}[h!]
    \centering
    \includegraphics[width=\linewidth]{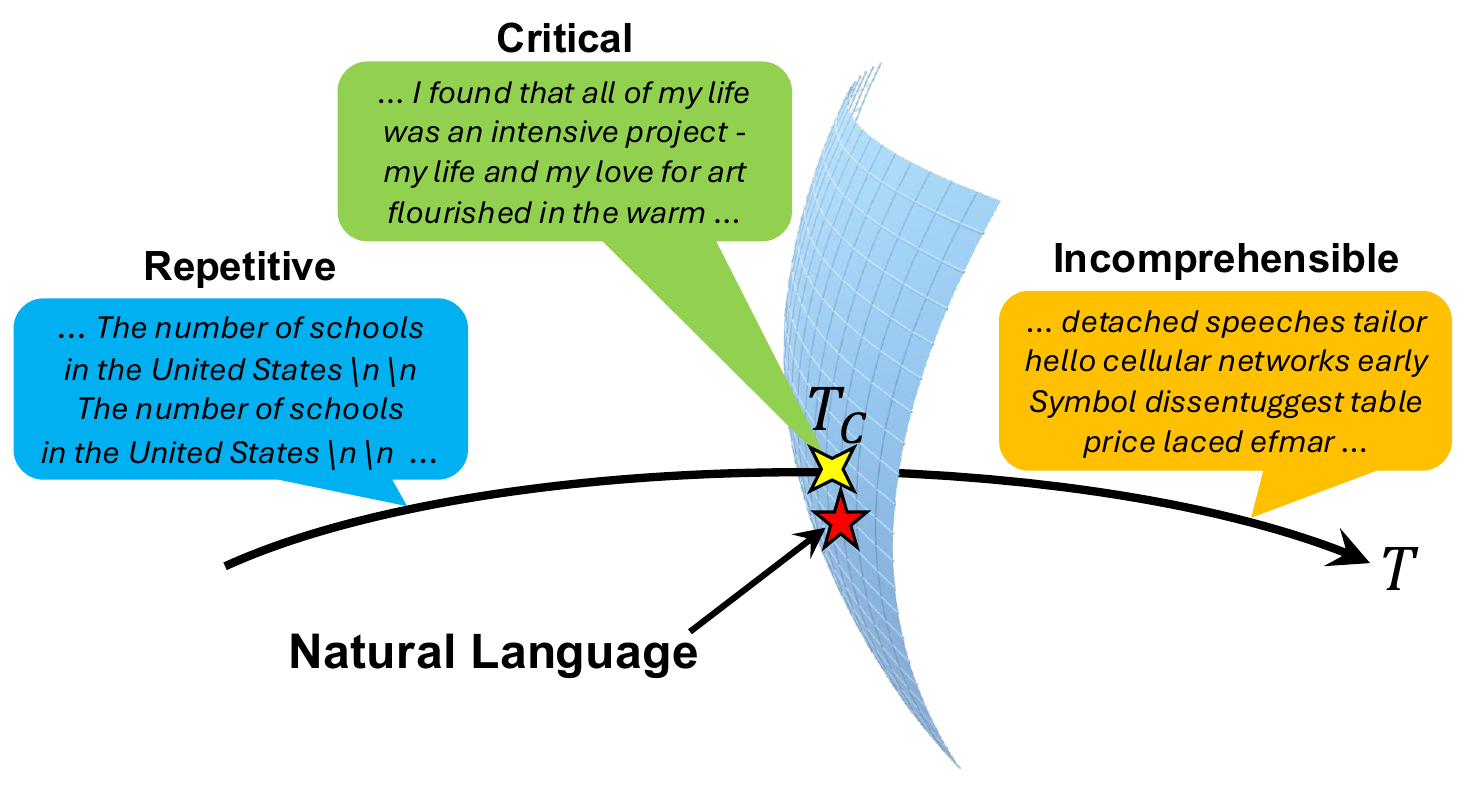}
    \caption{Schematic illustration of our conjecture. In a space of stochastic processes generating symbol sequences, the process corresponding to a natural language lies on the critical surface. An LLM trained on this natural language defines a one-dimensional manifold of stochastic processes parametrized by temperature $T$. At the critical temperature, this manifold intersects the critical surface near the point corresponding to the natural language.}
    \label{fig:intro}
\end{figure}

These findings strongly suggest that natural languages are indeed critical. Our conjecture is illustrated schematically in Fig.~\ref{fig:intro}. Consider a space in which each point corresponds to a stochastic process generating symbol sequences. The stochastic process corresponding to a natural language lies on a critical surface that separates this space into repetitive and incomprehensible phases. An LLM trained on this natural language defines a family of stochastic processes parameterized by the temperature parameter. This family of processes can be represented by a one-dimensional manifold in this space. At the critical temperature, this manifold intersects the critical surface, and this intersection lies near the point corresponding to the natural language. 

\section*{Phase transition in pretrained models}

To examine the conjecture that natural languages are critical, we use LLMs as effective models of natural languages. Specifically, we have generated texts using the 160M-parameter model from the Pythia suite~\cite{biderman2023pythia}, which is trained on the Pile, a large-scale English dataset~\cite{gao2020pile}. The generated texts were mapped to sequences of part-of-speech (POS) tags, which take 18 distinct values. Throughout this study, we treat the first $N$ tags $y_0, \cdots, y_{N-1}$ of each sequence as a time series of length $N$, with $y_t$ representing the state at time $t$. In Appendices~\ref{app:larger}--\ref{app:character}, we show that similar results are obtained for larger models, for models trained on different languages, and for another mapping of texts to sequences, supporting the robustness of the results presented in the main text.

\subsection*{Correlation}

In natural language texts, the choice of the next element, such as a word or character, depends on preceding elements through syntactic, semantic, and other linguistic factors. To quantify how far and how strongly such dependencies extend, one can examine how correlations between elements decay with distance. This approach has been widely adopted in previous studies, which have shown that the decay follows a power law~\cite{Li1989mutual, Ebeling1994-om, Ebeling1995-om, Tanaka-Ishii2016-oj, Lin2017-gs, takahashi2017neural, shen2019mutual, takahashi2019evaluating, Sainburg2019-tn, mikhaylovskiy2023autocorrelations}. This power-law behavior indicates that correlations can persist over remarkably long distances; in other words, distant elements in a text remain statistically dependent.

Motivated by these observations, we begin by quantifying correlations in texts generated by LLMs and examining their dependence on the temperature $T$. If a power-law decay is observed, it indicates that the generated texts reproduce a key statistical property of natural languages. Otherwise, it suggests a qualitative difference from natural languages. The time correlation between two states $y_t$ and $y_{t + \Delta t}$ separated by a time interval $\Delta t$ is defined by
\begin{equation}
    C_{ab}(t, t + \Delta t)
    = \mathbb{E} [
        \delta_{a, y_t} \delta_{b, y_{t + \Delta t}}
        ]
    - \mathbb{E} [ \delta_{a, y_t} ]
    \mathbb{E} [ \delta_{b, y_{t + \Delta t}} ],
\end{equation}
where $a$ and $b$ are POS tags such as NOUN and VERB, $\delta_{a,b}$ is the Kronecker delta, and $\mathbb{E} [ \cdot ]$ denotes an average over POS sequences generated at a fixed temperature $T$. A rapid decay of the correlation with $\Delta t$ indicates that texts are disordered. By contrast, if the correlation remains finite even at large $\Delta t$, texts have some order.

Among the $18^2$ possible pairs of $a$ and $b$, we mainly discuss the case in which both $a$ and $b$ are PROPN (proper noun), because this pair gives the largest contribution to the correlation. This choice is justified by the fact that the correlation for other pairs with large contributions shows similar behavior, as discussed in Appendix~\ref{app:integrated}. In what follows, we simply refer to $C_{\text{PROPN}, \text{PROPN}}$ as $C$. Unless otherwise specified, the same convention is used for other quantities.

\begin{figure}[t!]
    \centering
    \includegraphics[width=\linewidth]{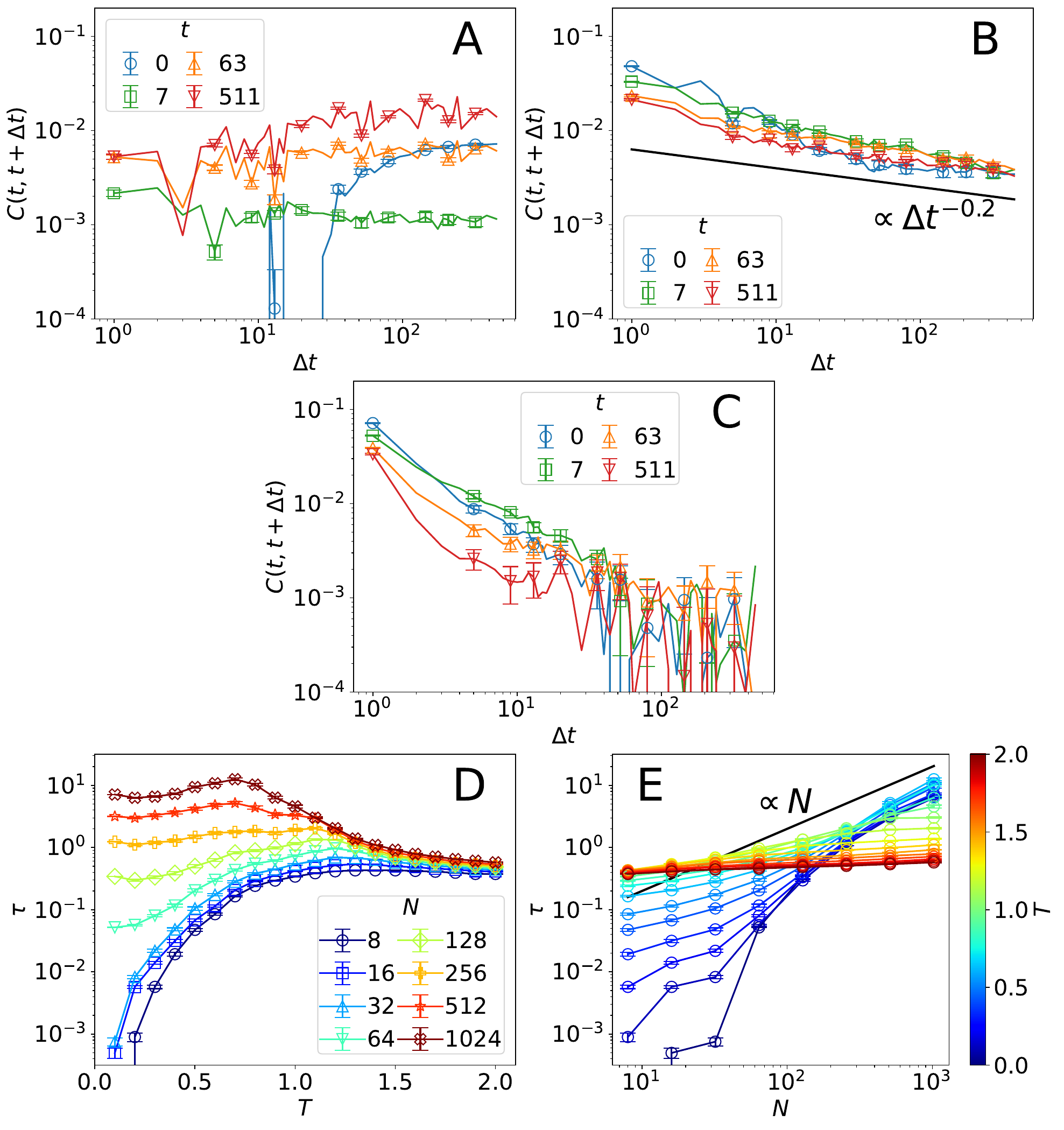}
    \caption{(A--C) Correlation $C(t, t+\Delta t) = C_{\text{PROPN},\text{PROPN}}(t, t+\Delta t)$ at (A) $T=0.4$, (B) $T=1$, and (C) $T=1.6$ as a function of time interval $\Delta t$, for sequence length $N = 1024$. The black line in (B) represents a line proportional to $\Delta t^{-0.2}$. The correlation saturates to a positive value at low temperatures and decreases to zero at high temperatures. Near the boundary temperature, the correlation follows a power-law decay. (D) Integrated correlation $\tau = \tau_{\text{PROPN}, {\text{PROPN}}}$ as a function of temperature $T$ for various sequence lengths $N$. (E) Same quantity as a function of sequence length $N$ for various temperatures $T$. The black line is proportional to $N$. The integrated correlation continues to increase with $N$ at $T \lesssim 1$, while it saturates at $T \gtrsim 1$. These results suggest that a phase transition occurs at $T_c \approx 1$.}
    \label{fig:pre_correlation}
\end{figure}

Figures~\ref{fig:pre_correlation}A--C show the correlation $C(t, t + \Delta t)$ at $T=0.4$, $1$, and $1.6$, respectively. At $T = 0.4$, the correlation converges to a positive value as the interval $\Delta t$ increases, whereas the plateau value tends to increase with the time $t$. This convergence to a positive value indicates that the system has long-range order. At $T = 1$, the correlation exhibits a power-law decay across different values of $t$, following $C \sim (\Delta t)^{-\eta}$ with $\eta \approx 0.2$. At $T = 1.6$, the correlation appears to be smaller for larger $t$, and its decay with $\Delta t$ is clearly faster than that at $T=1$. This behavior suggests that the system is disordered, although, within the present numerical accuracy, it remains unclear whether the correlation asymptotically follows an exponential decay, which is typically observed in a disordered phase.

To better understand the $T$-dependence of the correlation function, we compute the integrated correlation characterizing the time scale of the correlation:
\begin{align}
    \tau_{ab} &= \frac{1}{N} \sum_{t, t^\prime} C_{ab}(t, t^\prime) 
    = N \left(
        \mathbb{E} [ m_a m_b ] - \mathbb{E} [ m_a ] \mathbb{E} [ m_b ]
        \right),
    \label{eq:integrated_correlation}
\end{align}
where $m_a = \sum_t \delta_{a, y_t} / N$ is the proportion of POS tag $a$ in a sequence. If the correlation converges to a positive value, $\tau_{ab}$ diverges in the limit $N \to \infty$. By contrast, for short-range correlations, where the decay is faster than a power-law function, $\tau_{ab}$ remains finite in this limit. Thus, the $N$-dependence of $\tau_{ab}$ allows us to clearly discriminate between the two distinct behaviors. 

Figure~\ref{fig:pre_correlation}D presents the integrated correlation $\tau = \tau_{\text{PROPN}, \text{PROPN}}$ as a function of $T$. The $N$-dependence of the integrated correlation differs between low and high temperatures: At low temperatures, this quantity continues to increase with the sequence length $N$, whereas at high temperatures, it is almost insensitive to $N$. These distinct behaviors are more clearly seen in Fig.~\ref{fig:pre_correlation}E, which shows the same quantity as a function of $N$. In the low-temperature regime, $T \lesssim 1$, $\tau$ increases linearly with $N$, indicating that $\tau$ diverges in the limit $N \to \infty$. By contrast, at high temperatures $T \gtrsim 1$, it converges to a finite value, as expected for short-range correlations in a disordered phase. These observations indicate that $\tau$ becomes singular in the limit $N \to \infty$: As the temperature approaches a critical temperature $T_c \approx 1$ from above, the integrated correlation grows and eventually diverges at $T = T_c$. Such singular behavior establishes the presence of a phase transition at $T_c$.

The power-law decay of correlation indicates that the time scale of correlation diverges. Accordingly, the relaxation toward the stationary state is expected to be slow near $T_c$, a phenomenon known as \emph{critical slowing down}. In Appendix~\ref{app:dynamics}, we confirm this slowing down by observing how the probability that $y_t$ takes a given POS tag, $v_a(t) = \mathbb{E} [ \delta_{y_t, a} ]$, evolves with time $t$.

\subsection*{Power spectrum}

As shown in the previous section, the decay of correlation changes qualitatively at $T_c$. For $T < T_c$, the correlation converges to a positive value at large $\Delta t$, whereas for $T > T_c$, it decays rapidly. This behavior indicates that, at low temperatures, the generated texts exhibit an ordered structure that is absent at high temperatures. This observation raises the question of what kind of structure emerges below $T_c$. The repeated patterns observed in texts generated at low temperatures suggest that this structure is related to periodicity. We therefore analyze the power spectrum, a standard measure for detecting periodicity in a time series. This measure is defined by
\begin{equation}
    S_a (\omega)
    = N \left(
        \mathbb{E} \left[
            \left|
                f_a( \omega )
            \right|^2
        \right]
        - \left|
            \mathbb{E} \left[
                f_a( \omega )
            \right]
        \right|^2
    \right),
\end{equation}
where $f_a( \omega ) = \sum_t \mathrm{e}^{- 2 \pi i \omega t} \delta_{a, y_t} / N$ denotes the Fourier transform of the binary sequence $\delta_{a, y_t}$. A large peak in the power spectrum at frequency $\omega$ implies the presence of periodic structures with that frequency.

\begin{figure}[t!]
    \centering
    \includegraphics[width=\linewidth]{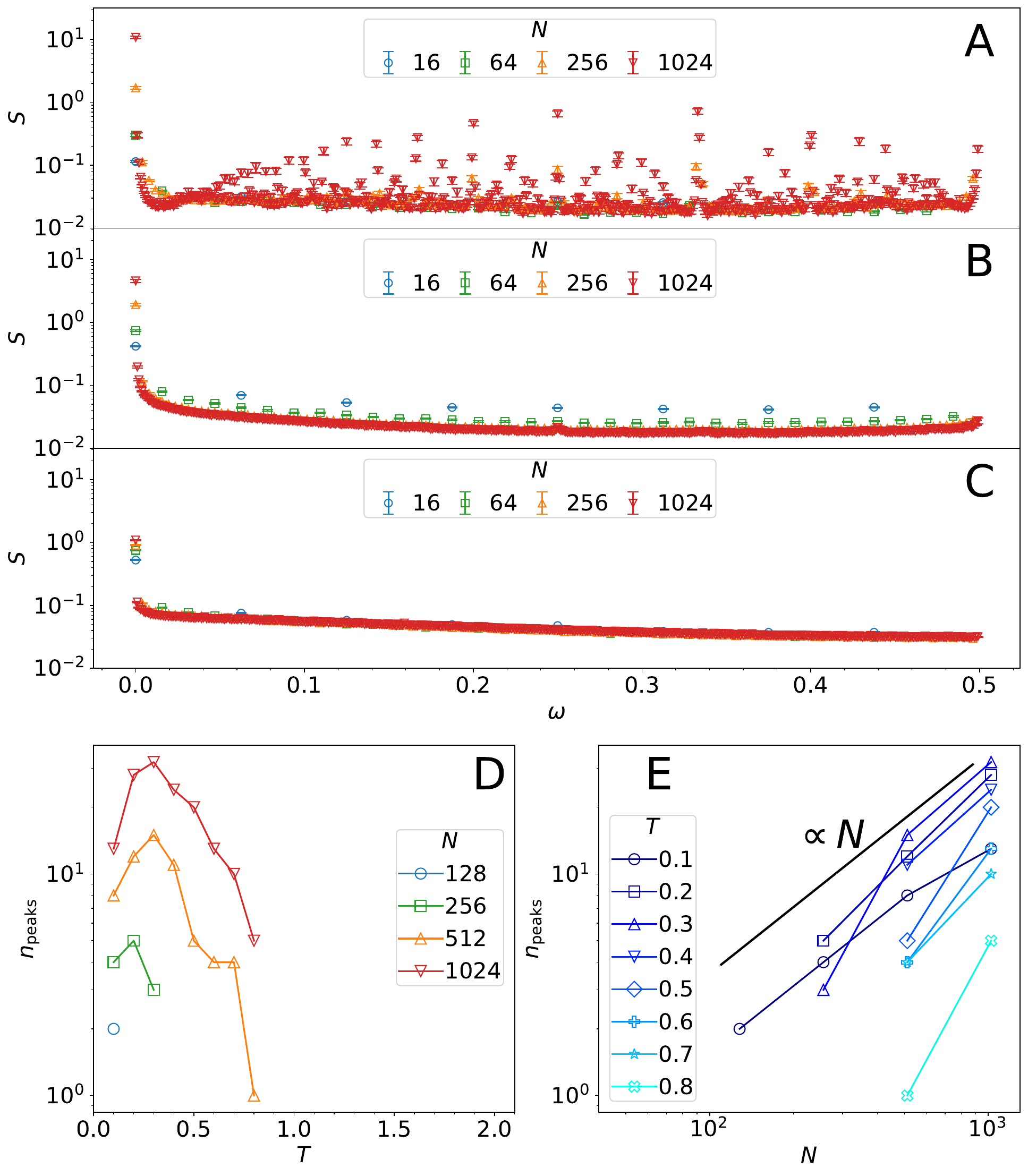}
    \caption{(A--C) Power spectrum $S = S_{\text{PROPN}}$ of POS sequences as a function of $\omega$ at (A) $T=0.6$, (B) $T=1$, and (C) $T=1.4$. (D) Number of peaks in the power spectrum, $n_{\text{peaks}}$, as a function of $T$ for various sequence lengths $N$. (E) Same quantity as a function of sequence length $N$ for various temperatures $T$. The black line represents a line proportional to $N$. $n_{\text{peaks}}$ diverges at low temperatures, indicating the complex repetitive structures.}
    \label{fig:pre_power}
\end{figure}

Figures.~\ref{fig:pre_power}A--C show $S(\omega) = S_{\text{PROPN}}(\omega)$ at $T=0.6$, $1$, and $1.4$, respectively. At $T = 1.4$, the spectrum has only a single, non-divergent peak at $\omega = 0$, consistent with a disordered structure in generated texts. At $T = 1$, very close to $T_c$, the spectrum exhibits multiple small peaks besides the one at $\omega = 0$, but their heights barely depend on $N$. As the temperature decreases, $S(\omega)$ develops many peaks with its heights growing with $N$, as observed at $T = 0.6$. These divergent peaks indicate the presence of inhomogeneous long-range order. This spectral behavior is consistent with empirical observations. At low temperatures, generated texts typically fall into repetitions such as \emph{The number of schools in the United States\textbackslash{}n\textbackslash{}nThe number of schools in the United States\textbackslash{}n\textbackslash{}n...}. As confirmed by analyses of the power spectrum defined for individual sequences, such a repetitive text yields peaks at $\omega = n/p$ ($n = 1, 2, \cdots$), where $p$ is the period of the repetition  (see Appendix~\ref{app:power} for detail). Averaging over many texts, each with a distinct period, gives rise to a power spectrum with many peaks. 

Our analysis reveals an even richer structure at low temperatures. Figures~\ref{fig:pre_power}D and E show the number of peaks in the power spectrum, $n_{\text{peaks}}$, as a function of $T$ and $N$, respectively. At $T < T_c$, $n_{\text{peaks}}$ grows linearly with $N$, indicating that it diverges as $N \to \infty$. This $N$-dependence suggests that the repeated patterns become more diverse as the text length increases and that infinitely many distinct patterns emerge at $N \to \infty$. This behavior contrasts sharply with the simple periodic structures such as those observed in limit cycles in dynamical systems~\cite{berge1986order} and periodic Markov chains~\cite{norris1997markov}.

\section*{Emergence of the phase transition during training}

In the previous section, we showed that texts generated by the pretrained model exhibit two notable phenomena: the critical behavior at $T_c$ and the presence of complex repetitive structures at low temperatures. These phenomena are expected to emerge through training on the natural language dataset. We therefore systematically analyze texts generated by partially trained models at different training steps. The results show that both phenomena are absent at early training steps but emerge at later steps, suggesting that they originate from natural languages.

\subsection*{Correlation}

\begin{figure}[t!]
    \centering
    \includegraphics[width=\linewidth]{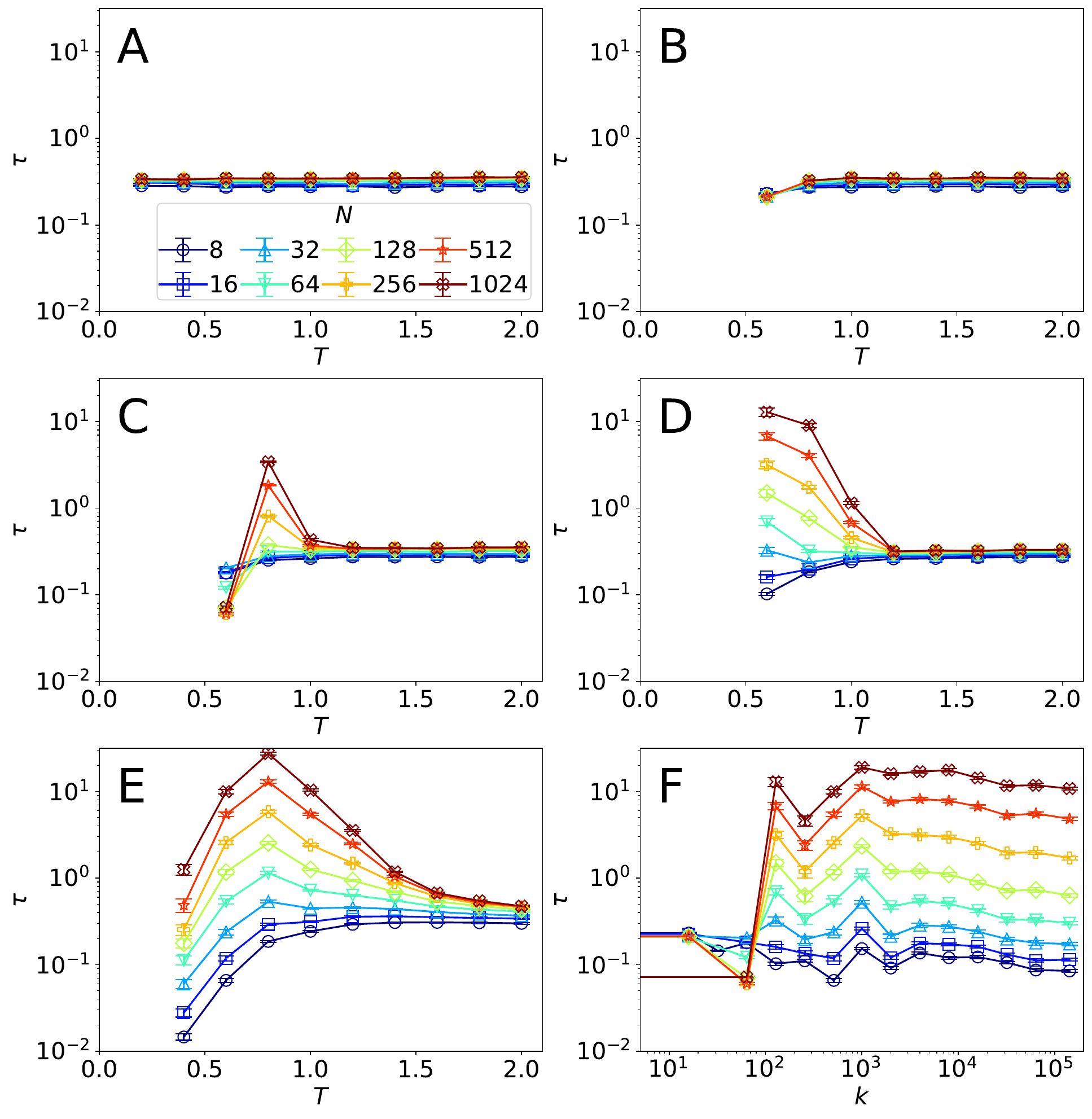}
    \caption{(A--E) Integrated correlation $\tau = \tau_{\text{PROPN}, {\text{PROPN}}}$ as a function of temperature $T$ at training steps $k = 0$ (A), 16 (B), 64 (C), 128 (D), and 512 (E). (F) Integrated correlation as a function of training step $k$ at $T = 0.6$. These results suggest that the phase transition emerges around $k_c \approx 10^2$. Several values of the integrated correlation at early training steps and low temperatures are not shown because text generation terminates too early to obtain a sufficient number of long sequences.}
    \label{fig:par_corr}
\end{figure}

Since the integrated correlation $\tau = \tau_{\text{PROPN}, \text{PROPN}}$ provides clear evidence for the phase transition in the pretrained model, we use the same quantity to explore how generated texts change over training steps. In Figs.~\ref{fig:par_corr}A--E, we show $\tau$ at training step $k = 0$, 16, 64, 128, and 512. At $k = 0$ and 16, $\tau$ remains nearly constant across temperatures, indicating that the phase transition observed in the pretrained model is absent in the very early stage of training, as expected. The dependence of $\tau$ on $T$ and $N$ gradually changes during the training, and at $k \gtrsim 128$, the dependence becomes very similar to that of the pretrained model. Figure~\ref{fig:par_corr}F shows the $k$-dependence of $\tau$ at a low temperature, $T = 0.6$. The integrated correlation is almost independent of $N$ for $k \lesssim 10^2$, but suddenly starts to grow with $N$ at $k = 128$. These results indicate that the model begins to acquire nontrivial structures of the natural language around $k_c \approx 10^2$, giving rise to a critical phase transition. As discussed in Appendix~\ref{app:slowing}, critical slowing down in the mean time series $v_a (t)$ also appears at $k \gtrsim 10^2$, consistent with the results on $\tau$.

\subsection*{Power spectrum}

\begin{figure}[t!]
    \centering
    \includegraphics[width=\linewidth]{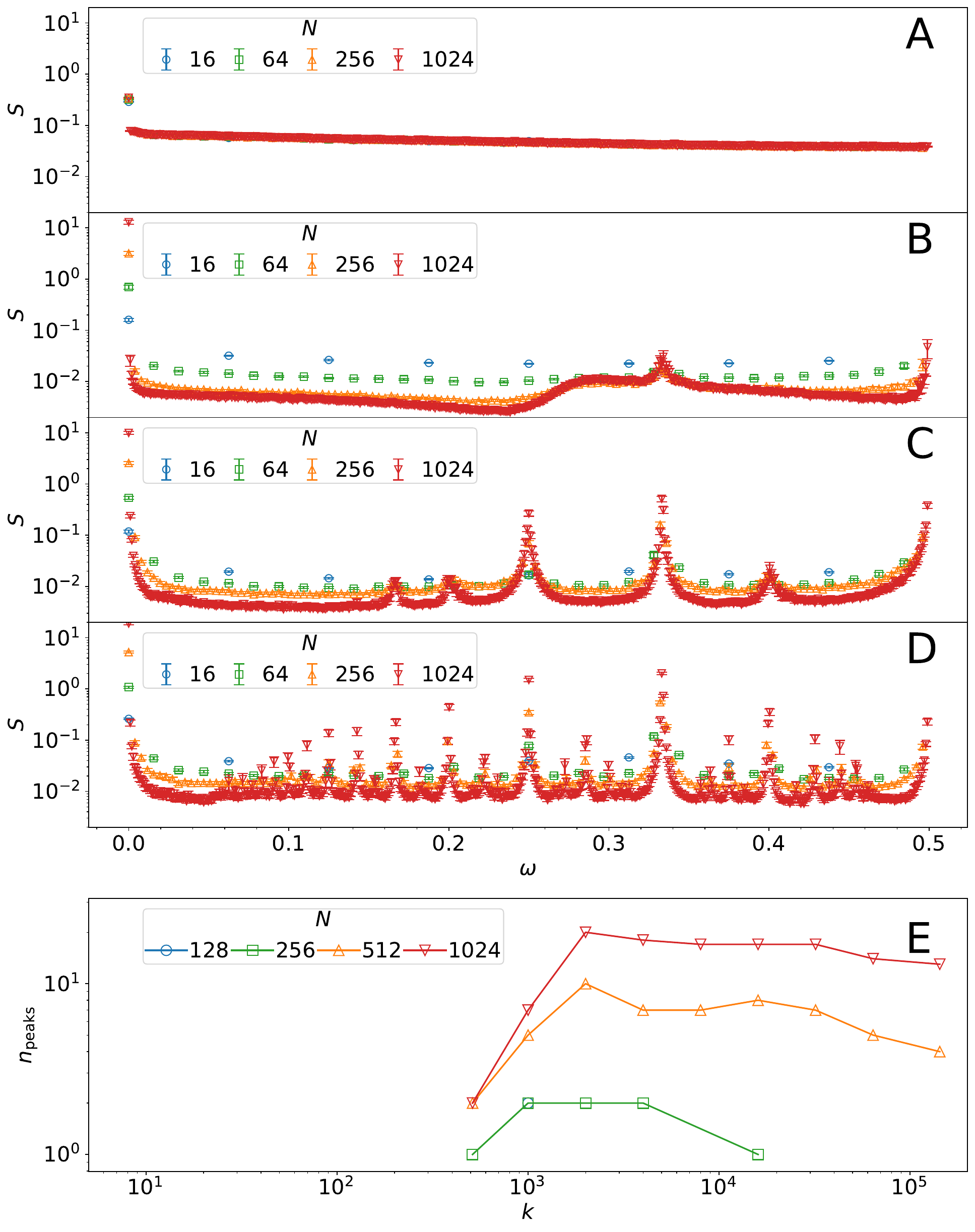}
    \caption{(A--D) Power spectrum $S = S_{\text{PROPN}}$ for $T=0.6$ at training step  (A) $k = 0$, (B) 128, (C) 512, and (D) 1000. (E) Number of peaks, $n_{\text{peaks}}$, in the power spectrum at $T=0.6$ as a function of training step. $n_{\text{peaks}}$ begins to grow around training step $k \approx 10^3$, which is later than the step at which the phase transition emerges.}
    \label{fig:par_power}
\end{figure}

To examine how complex repetitive structures at low temperatures emerge during training, we measure the power spectrum at different training steps, as is done for the pretrained model. In the high-temperature regime, the power spectrum does not exhibit any notable features at any training steps: It has only a single peak at $\omega = 0$, consistent with the fact that texts generated at high temperatures do not display any nontrivial structure at any stage of training.
In the low-temperature regime, by contrast, training has a significant influence on the power spectrum, see Figs.~\ref{fig:par_power}A--D. For the randomly initialized model, i.e., the model at training step $k = 0$, the power spectrum exhibits only a single peak at $\omega = 0$, indicating the absence of any repetitive structure. At step $k=128$, multiple peaks apart from the one at $\omega = 0$ start to appear at $\omega = 1/2$ and $1/3$, yet their heights are almost independent of the sequence length $N$. These observations are consistent with our observation that period-2 and period-3 repetitions are present but not dominant, and most of the generated texts have period-1 repetitions such as \emph{""" ...} and \emph{the the the ...}. At $k = 512$, $S(\omega)$ develops a few peaks with their heights growing with $N$. With increasing $k$ further, the power spectrum becomes more complex, and at $k \gtrsim 10^3$, it displays numerous growing peaks at $\omega \neq 0$, as observed in the pretrained model. The number of peaks, $n_{\text{peaks}}$, begins to grow with $N$ only around $k \approx 10^3$, see Fig.~\ref{fig:par_power}E. This onset occurs later than the training step $k_c \approx 10^2$, at which the integrated correlation begins to show critical behavior. The difference between these two onsets suggests that the model acquires two distinct structures at different stages of training: one corresponding to criticality and the other to complex repetitive structures.

\section*{Natural language is closest to the critical model}

The results presented above demonstrate that the pretrained model near the critical temperature $T_c$ exhibits power-law behavior, which has also been observed in natural languages. In Appendix~\ref{app:natural}, we further confirm that the statistical behavior of natural languages is similar to that of texts generated by the critical pretrained model in terms of the correlation, integrated correlation, mean time series, and power spectrum. From these results, we expect that, among the models at different training steps and temperatures, the one at the final training step and near $T_c$ generate texts closest to natural language texts.

\begin{figure}[h!]
    \centering
    \includegraphics[width=.8\linewidth]{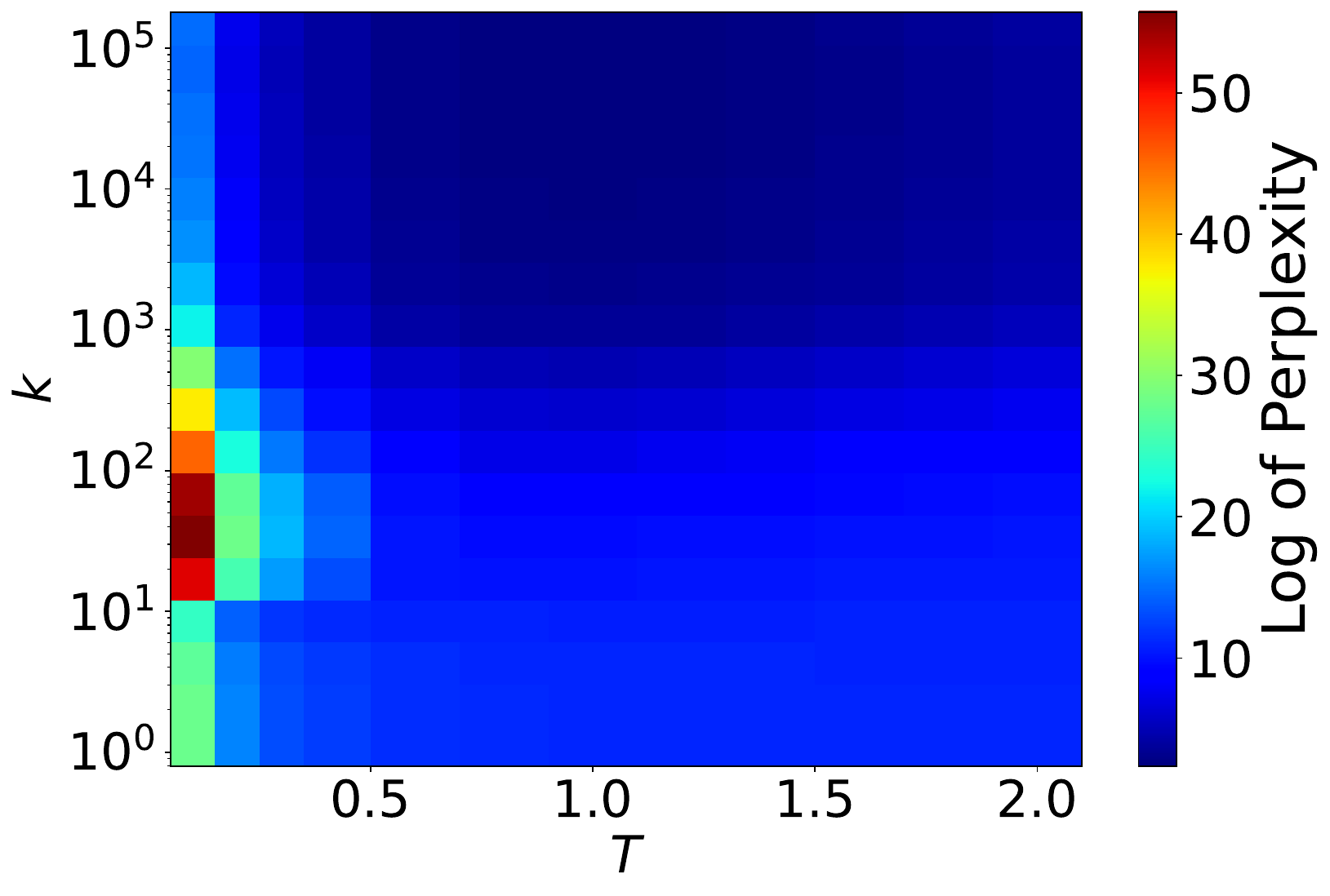}
    \caption{Natural logarithm of perplexity of the models on the Pile at different temperatures and training steps. Perplexity is lowest for the pretrained model around the critical point, supporting the conjecture that natural languages are critical. At low temperatures, perplexity is highest around training step $k = 32$, consistent with the model distribution concentrating on simple, repetitive texts in this regime.}
    \label{fig:natural}
\end{figure}

To examine this expectation quantitatively, we measure the agreement between natural language texts and texts generated by models at different training steps and temperatures. Specifically, we use perplexity, a widely used metric for evaluating language models in natural language processing~\cite{jurafsky2026speech}. This metric is defined as the exponential of the average negative log-likelihood of the model for a given natural language dataset. Lower perplexity therefore indicates better agreement between the stochastic process defined by the model and the natural language dataset. Figure~\ref{fig:natural} shows the natural logarithm of perplexity on the Pile. The perplexity is lowest at $k = 143000$, the final training step, and at $T = 1$, close to $T_c$. This result confirms quantitatively that the critical pretrained model most closely matches natural language texts in the dataset. This agreement provides evidence for the conjecture that natural languages are critical, as illustrated in Fig.~\ref{fig:intro}.

We also note that, at low temperatures, perplexity is nonmonotonic as a function of $k$; it becomes remarkably large around $k=32$. This nonmonotonic behavior can be understood from changes in the generated texts during training. At very small $k$, the texts remain disordered even at low temperatures because the model has not yet acquired nontrivial structures. Around $k=32$, the model distribution at low temperatures concentrates on a limited number of extremely simple repetitive texts such as \textit{\textbackslash{}n\textbackslash{}n\textbackslash{}n ...} and \textit{the the the ...}. The high perplexity around this point indicates that such a distribution is farther from the natural language dataset than is the nearly uniform distribution in the very early stage of training. As training proceeds further, repetitions persist, yet the repeated patterns become more natural and diverse, as in \textit{You can use a button.\textbackslash{}n\textbackslash{}nYou can use a button.\textbackslash{}n\textbackslash{}n ...} and \textit{I have a new value of \$50\textbackslash{}n\textbackslash{}nI have a new value of \$50\textbackslash{}n\textbackslash{}n ...}. This change is consistent with the decrease in perplexity, indicating that the model distribution becomes closer to the natural language dataset.

\section*{Summary and Discussion}

Natural languages exhibit power-law behavior in their statistical properties. This well-known fact suggests a connection between natural languages and critical phenomena. We have examined this potential connection using LLMs as effective models of natural languages. Our analyses of LLM-generated texts show that they undergo a phase transition at $T_c \approx 1$ with critical behavior, and that the low-temperature phase exhibits complex repetitive structures with a divergent number of peaks in the power spectrum, in sharp contrast to simple periodic structures with a finite number of peaks. The critical behavior and the complex repetitive structures originate from natural languages, because both are absent in the early stage of training but emerge as the model learns from the natural language dataset. Finally, we numerically show that the natural language dataset is statistically closest to texts generated by the critical pretrained model as measured by perplexity. As discussed in Appendices~\ref{app:larger} and \ref{app:other}, similar results are obtained for larger models and for models trained on other languages. These additional results suggest that the phenomena observed here are robust across model sizes and languages.

Taken together, our analyses provide strong evidence for the conjecture that natural languages are critical: The stochastic process corresponding to a natural language lies on a critical surface separating the repetitive and incomprehensible phases. In the former phase, the number of peaks in the power spectrum diverges as the text length increases. Such behavior is not observed in the power spectra of natural languages. Nevertheless, since it emerges as a result of the training of the model on a natural language, this behavior originates from some structure in the natural language. Identifying this structure is an important open problem.

The present work opens a route to studying natural languages using concepts developed in the study of critical phenomena. One such concept is that of \emph{universality classes}. If the critical behaviors of different systems are characterized by the same set of exponents, the systems are classified into the same universality class. Systems in the same class share fundamental features such as symmetries, dimensionality, and interaction range. Such exponents can be estimated for LLMs trained on different natural languages. One possible direction is to classify natural languages according to these exponents and ask what fundamental features are shared by languages in the same class. This approach may provide a framework for characterizing natural languages in terms of their macroscopic behavior.

Our results have implications not only for natural languages but also for LLMs. A future direction for studying LLMs concerns the relationship between their performance and their critical behavior. Generated texts are expected to remain neither repetitive nor incomprehensible within the time scale over which the power-law behavior persists. This scale is finite at temperatures away from the critical point and diverges exactly at the point. If the temperature dependence of this scale could be predicted, it would provide an upper bound on the sequence length beyond which the model fails to generate natural texts. This perspective may also help explain why some modern LLMs rarely produce repetitions even when the temperature is substantially lower than $T_c$. For models with large context windows and various heuristics, this scale is likely to be sufficiently longer than the sequence lengths required in practical tasks, even at low temperatures.

Our work builds on the view of natural languages as stochastic processes and is motivated by the power-law behavior observed in natural languages. This view has a long history, and the power-law behavior in natural languages has long been studied. The key idea of our work is to use LLMs as effective models of natural languages. This approach allows us to argue that natural languages are critical, a conclusion that was difficult to obtain using toy mathematical models. We believe that our work provides a concrete step toward understanding the nature of natural languages from a statistical physics perspective.

\section*{Materials and Methods}

\subsection*{Temperature parameter in large language models}

LLMs autoregressively generate a sequence of tokens, which correspond to words or subwords, and then convert it into a text. In temperature sampling \cite{ackley1985learning}, the $t$-th token $x_t$ is drawn from the softmax distribution
\begin{equation}
    \label{eq:temp_samp}
    P ( x_t | x_0, \cdots, x_{t-1} )
    \propto \exp \left( 
        - \frac{1}{T} H(x_t | x_0, \cdots, x_{t-1})
    \right),
\end{equation}
where $T$ is the temperature parameter and $H$ denotes the negative logit associated with the candidate token $x_t$ given the preceding tokens. When $T$ is set to its default value of 1, the softmax distribution coincides with the learned distribution without modification. If the temperature is lower, the distribution becomes more biased to high-probability tokens, whereas with increasing temperature, the distribution approaches a uniform one. This parameter is analogous to physical temperature, as the functional form of Eq.~\ref{eq:temp_samp} formally corresponds to the Boltzmann distribution in statistical physics. However, we stress that Eq.~\ref{eq:temp_samp} determines the probabilities of individual tokens conditioned on the tokens generated in the past, in contrast to the Boltzmann distribution, which gives the time-independent probabilities of states.

\subsection*{Sampling method}

We used models from the Pythia suite \cite{biderman2023pythia}. This suite provides models of various sizes and at different training steps on the same English dataset, enabling experiments under controlled conditions. All the data presented in the main text were generated using the model with 160M parameters. Additional analyses to check the robustness of our results were performed on larger models with 410M, 1B, 1.4B, and 2.8B parameters, as well as on models trained on other languages, that is, Chinese, French, Japanese, Korean, and Spanish, see Appendix~\ref{app:larger} and \ref{app:other}.

For each sequence length $N$ and temperature $T$, we sampled $10^4$ POS sequences for all models, except for the correlation functions in Figs.~\ref{fig:pre_correlation}A--C, for which we sampled $1.6 \times 10^5$ sequences to ensure reliable estimation at large distances. Texts were generated using the beginning-of-sentence token as input. We did not use top-$k$ sampling, top-$p$ sampling, or beam search, in order to preserve the analogy between Eq.~\ref{eq:temp_samp} and the Boltzmann distribution. Error bars in the figures represent $80\%$ confidence intervals estimated using the symmetric bootstrap-$t$ method~\cite{hall1988symmetric}.

\subsection*{Mapping to part-of-speech sequences}

To conduct statistical analyses, we need to map each text to a sequence of a finite number of states. In this study, we represented each text as a sequence of POS tags using the spaCy library~\cite{montani_2023_10009823} with the en\_core\_web\_sm pipeline, where one of 18 tags (17 universal POS tags plus a SPACE tag) is assigned to each word or space. For instance, the sentence \emph{The first time I saw the film.} is transformed into $(\text{DET}, \text{ADJ}, \text{NOUN}, \text{PRON}, \text{VERB}, \text{DET}, \text{NOUN}, \text{PUNCT})$. 
In this setup, each state admits a clearer interpretation and retains more linguistic information than alternatives such as mapping each character to a number~\cite{Li1989mutual, Ebeling1995-om}. Moreover, the moderate number of possible states enables reliable estimation of statistical quantities. Finally, texts in different languages can be mapped to sequences of the same tags, enabling consistent cross-linguistic analysis, as conducted in Appendix~\ref{app:other}.

At very low or high temperatures, generated texts sometimes contain unnatural subsequences, yet the tagger automatically assigns POS tags. Although this process could in principle introduce artifacts, such an effect should be negligible near the critical point, which is our main focus. Indeed, we did not observe any behavior suggesting a significant effect. To further support that our conclusions are not artifacts of POS tagging, we also conducted the same analyses under an alternative setup that does not involve such effects. In this setup, each text was converted into a binary sequence by mapping each character to 0 or 1, depending on whether it was a space or not. The results are similar to those obtained with POS tagging, as presented in Appendix~\ref{app:character}.

To study POS sequences of length $N$, we extracted sequences that are longer than $N$ tags and analyzed the first $N$ tags, $y_0, \cdots, y_{N-1}$, from the beginning of each sequence.

\subsection*{Peak detection in the power spectrum}

We counted the number of peaks in the power spectrum, $n_{\text{peaks}}$, using the find\_peaks function from the SciPy library \cite{virtanen2020scipy}. Specifically, we identified the power spectrum $S(\omega)$ as having a peak at frequency $\omega$ when the prominence at $\omega$ is greater than a threshold. The prominence is defined by $S (\omega) - \min_{\omega_{\text{left}} < \omega^\prime < \omega_{\text{right}}} S(\omega^\prime)$, where $\omega_{\mathrm{left}}$ and $\omega_{\mathrm{right}}$ are the frequencies $\omega^\prime$ closest to $\omega$ such that $S(\omega^\prime) > S(\omega)$. We chose a threshold of $0.1$ to exceed the estimated statistical error.

\subsection*{Computation of perplexity}

Computing perplexity exactly is impractical because it requires evaluating the likelihood for every token conditioned on the entire preceding context. In practice, perplexity is approximated by sliding a window of the same size as the model’s context length across the dataset with a fixed stride~\cite{huggingface2025ppl}. In this study, we computed perplexity on $200$ texts from the Pile using a stride of $512$ tokens.

\subsection*{Data, Materials, and Software Availability}

The source code is available at \url{https://github.com/K-Nakaishi/transition-in-llm}~\cite{nakaishi2026transition}.

\begin{acknowledgments}
We thank Y.~Ichikawa, H.~Ikeda, J.~Takahashi, T.~Takahashi, N.~Mikhaylovskiy, J.~Arnold, N.~L\"{o}rch, and S.~Yokoi for useful discussions and suggestions. This work was supported by JSPS KAKENHI Grant Nos.~22K13968, 23KJ0622, 23H01095, and 25K24434, JST Grant Number JPMJPF2221, and the World-Leading Innovative Graduate Study Program for Advanced Basic Science Course at the University of Tokyo.
\end{acknowledgments}


\appendix

\section{Larger models}
\label{app:larger}

In the main text, we analyzed texts generated by the 160M-parameter model in the Pythia suite~\cite{biderman2023pythia}. However, much larger models are used in practice. To confirm that larger models exhibit similar phenomena, we estimate the integrated correlation and the power spectrum for Pythia 410M, 1B, 1.4B, and 2.8B using the same setup.

\begin{figure*}[t]
    \centering
    \includegraphics[width=.76\linewidth]{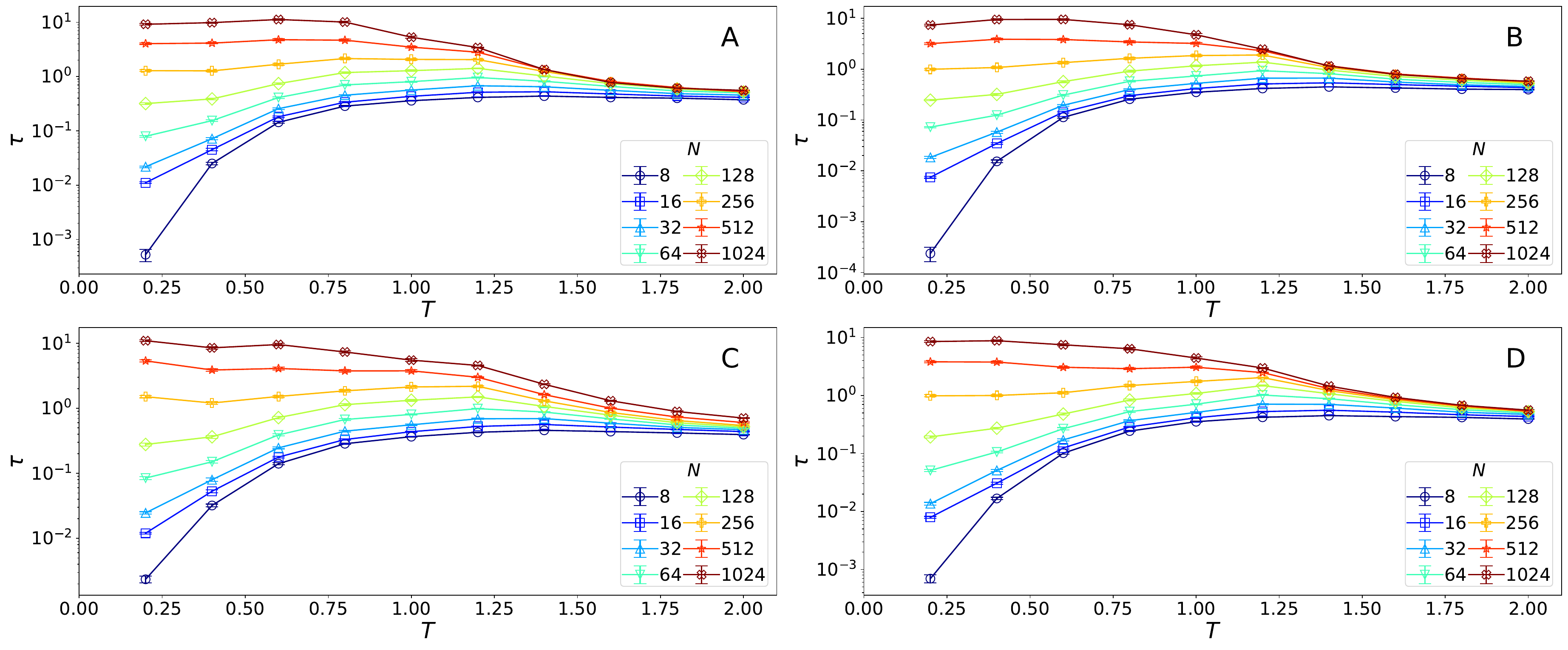}
    \caption{Integrated correlation $\tau = \tau_{\text{PROPN}, \text{PROPN}}$ for texts generated by Pythia (A) 410M, (B) 1B, (C) 1.4B, and (D) 2.8B. The phase transition is consistently observed across all model sizes.}
    \label{fig:pnas_SI_larger_corr}
\end{figure*}

\begin{figure*}[t]
    \centering
    \includegraphics[width=.76\linewidth]{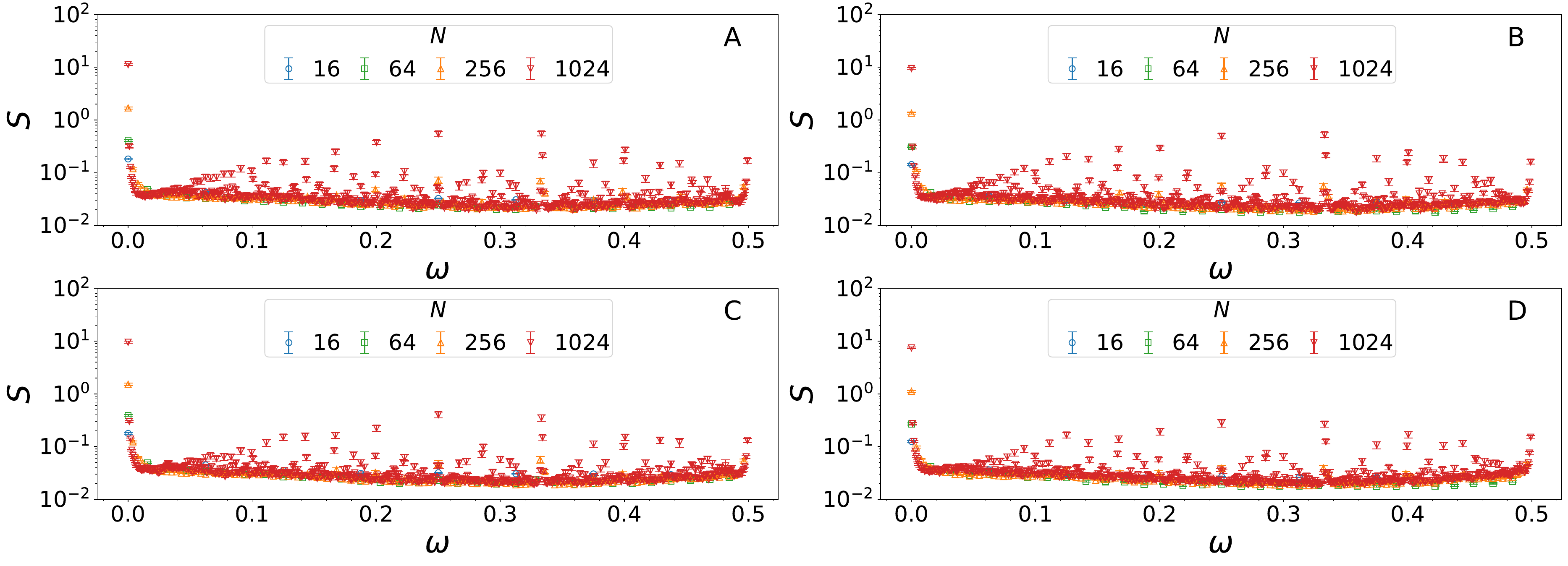}
    \caption{Power spectra for texts generated at $T = 0.6$ by Pythia (A) 410M, (B) 1B, (C) 1.4B, and (D) 2.8B. As observed for Pythia 160M, numerous peaks appear, indicating that the generated texts exhibit the complex repetitive structure.}
    \label{fig:pnas_SI_larger_power}
\end{figure*}

The integrated correlation $\tau = \tau_{\text{PROPN}, {\text{PROPN}}}$ for the larger models, shown in Fig.~\ref{fig:pnas_SI_larger_corr}, exhibits behavior similar to that observed for Pythia 160M: $\tau$ diverges at low temperatures and saturates at high temperatures. The power spectrum $S(\omega) = S_{\text{PROPN}}(\omega)$ shows a similar correspondence. At high temperatures, the spectrum has only a single peak at $\omega = 0$, indicating that the generated texts have a trivial structure. By contrast, numerous divergent peaks appear at low temperatures, as in the results for $T = 0.6$ shown in Fig.~\ref{fig:pnas_SI_larger_power}. These results indicate that both the phase transition and the complex repetitive structures are present across different model sizes.

In the main text, we also analyzed the perplexity of partially trained 160M-parameter models. The results show that the perplexity is lowest for the critical pretrained model and highest near training step $k = 32$ at low temperatures. To examine whether these results are robust to model size, we compute the perplexity of partially trained models with 410M, 1B, 1.4B, and 2.8B parameters.

\begin{figure*}[t]
    \centering
    \includegraphics[width=.57\linewidth]{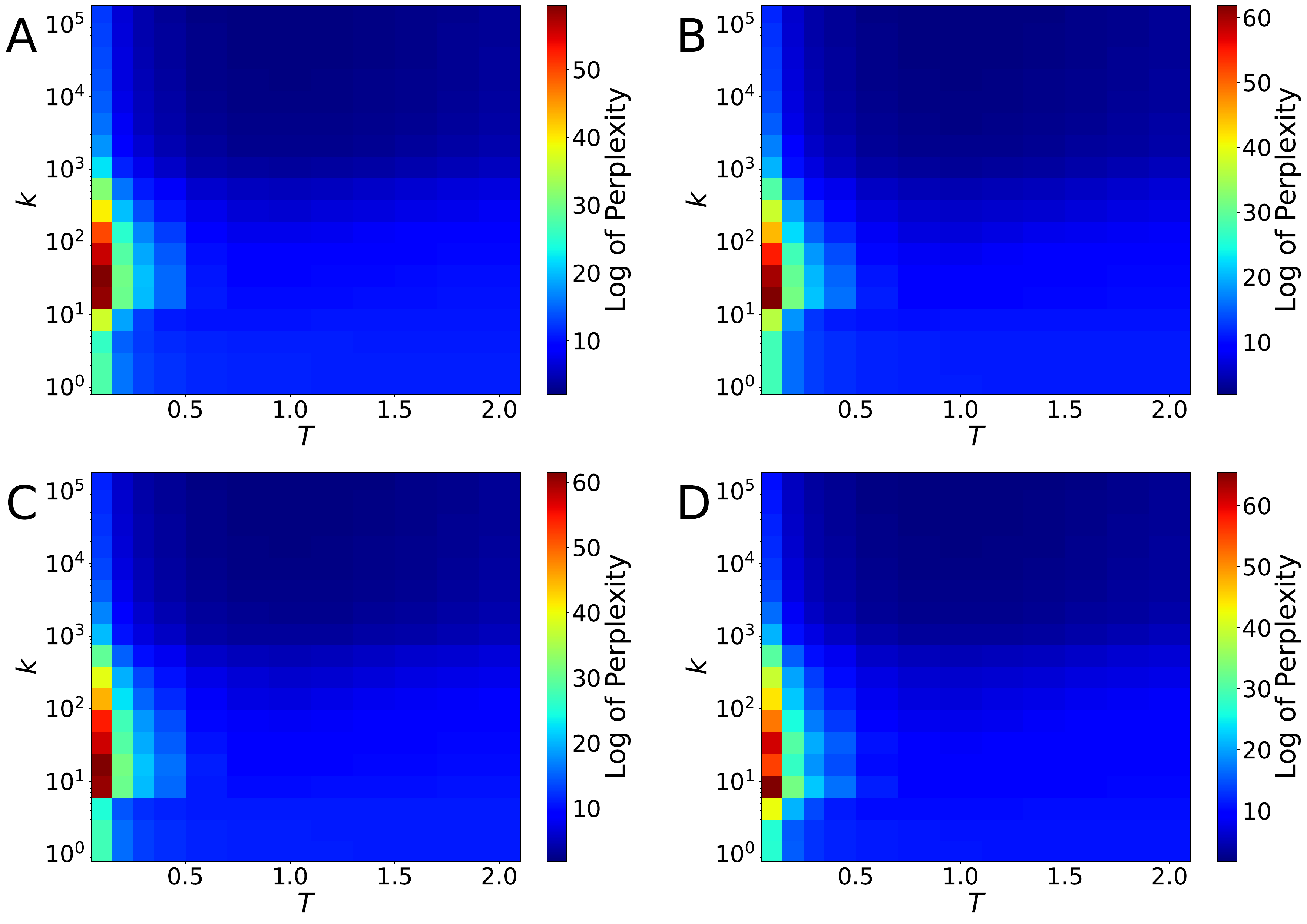}
    \caption{Natural logarithm of perplexity of Pythia (A) 410M, (B) 1B, (C) 1.4B, and (D) 2.8B on the Pile at different temperatures $T$ and training steps $k$. For all model sizes, it is lowest for the critical pretrained model and remarkably high near step $k = 32$ at low temperatures}
    \label{fig:pnas_SI_perplexity}
\end{figure*}

The results are shown in Fig.~\ref{fig:pnas_SI_perplexity}. The behavior of perplexity is consistent across model sizes. It is minimized for the critical pretrained model, indicating that the generated texts are closest to the natural language dataset near this point. In addition, the perplexity becomes markedly high near step $k = 32$ at low temperatures. These results suggest that the statistical properties of generated texts evolve in a similar way across model sizes.

\section{Other languages}
\label{app:other}

The Pythia suite, discussed in the main text, is trained on the English dataset. A natural question is whether the observed phenomena also occur for models trained on other languages. To address this question, we perform the same analyses on models trained on Chinese, French, Japanese, Korean, and Spanish datasets. Specifically, we analyze uer/gpt2-chinese-cluecorpussmall~\cite{uer2021gpt2_chinese}, dbddv01/gpt2-french-small~\cite{dbddv2020french}, rinna/japanese-gpt2-small~\cite{zhao2021japanese, sawada2024release}, SKT-AI/KoGPT2~\cite{skt2021kogpt}, and datificate/gpt2-small-spanish~\cite{datificate2021spanish}, using the same setup as in the main text.

\begin{figure*}[h]
    \centering
    \includegraphics[width=\linewidth]{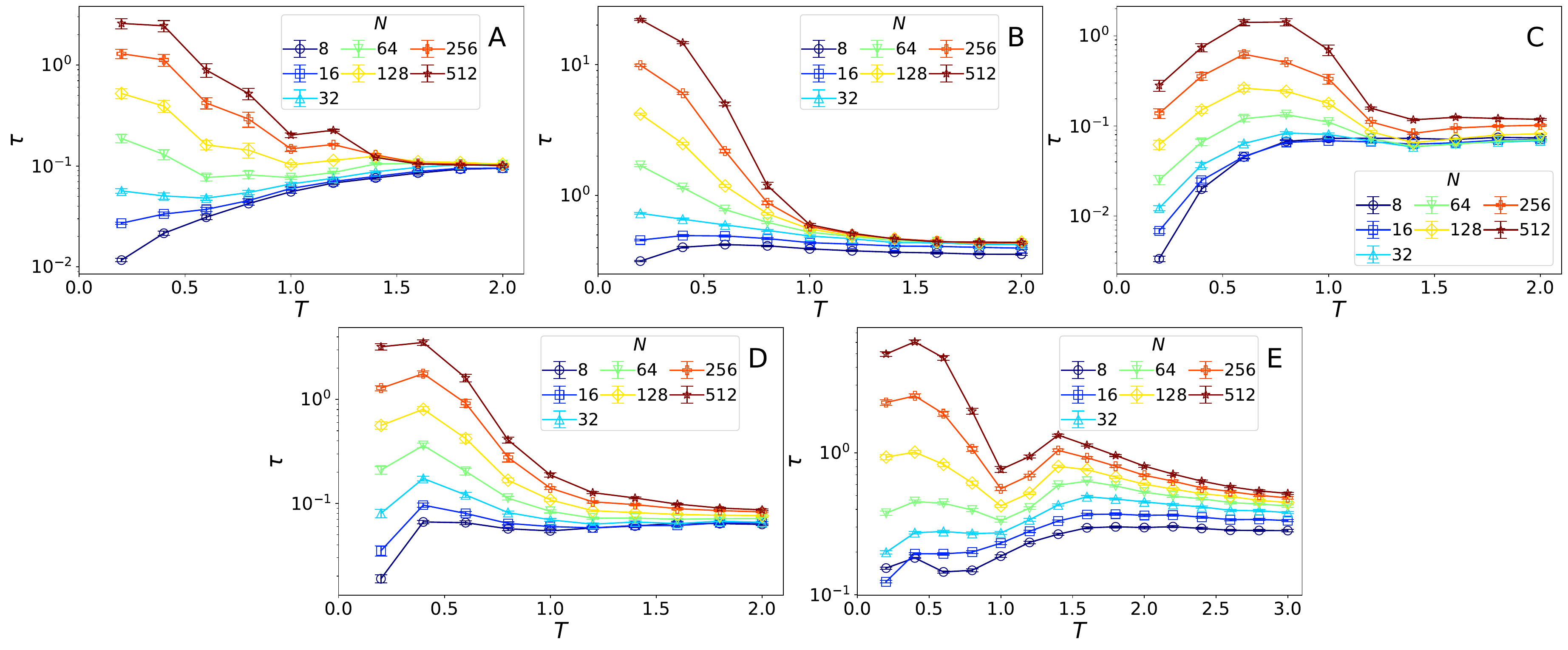}
    \caption{Integrated correlation $\tau = \tau_{\text{PROPN},\text{PROPN}}$ for texts generated by models trained on (A) Chinese, (B) French, (C) Japanese, (D) Korean, and (E) Spanish. The phase transition is consistently observed across different languages.}
    \label{fig:pnas_SI_multiling_corr}
\end{figure*}

\begin{figure*}[h]
    \centering
    \includegraphics[width=\linewidth]{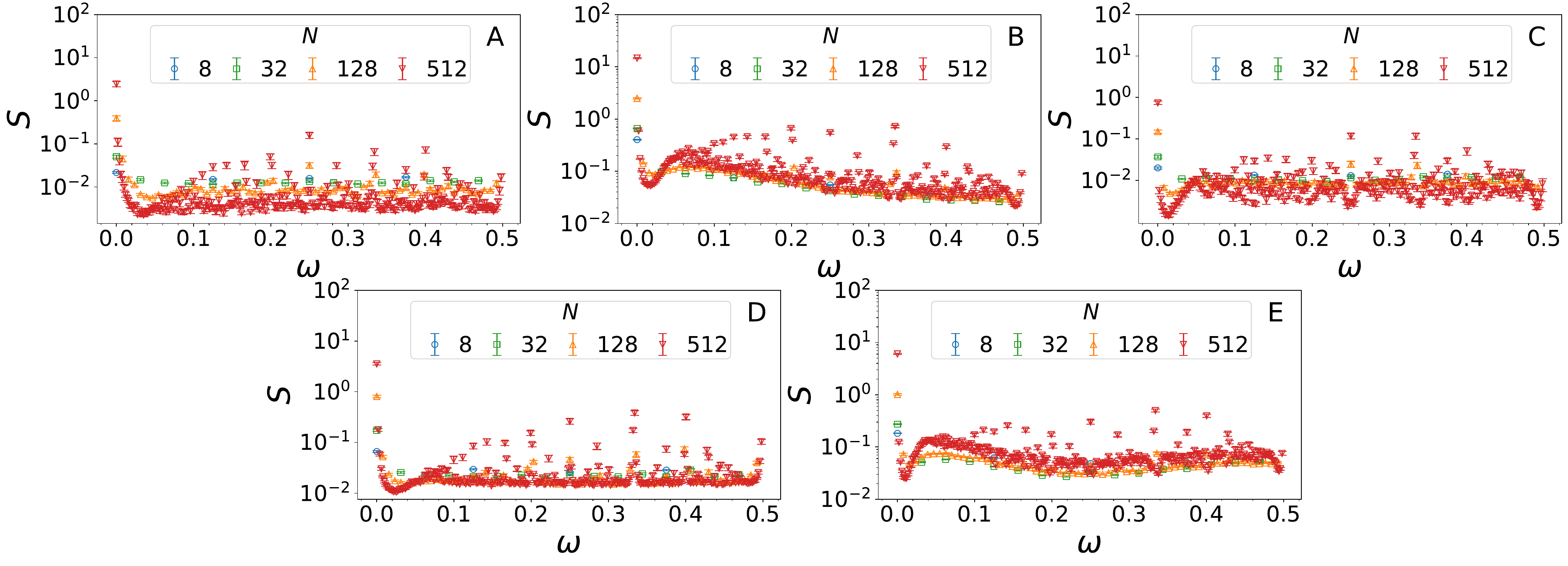}
    \caption{Power spectra $S = S_{\text{PROPN}}$ for texts generated at $T = 0.4$ by models trained on (A) Chinese, (B) French, (C) Japanese, (D) Korean, and (E) Spanish. As for Pythia trained on English, numerous peaks appear, indicating that generated texts have the complex repetitive structures.}
    \label{fig:pnas_SI_multiling_power}
\end{figure*}

Across these languages, the integrated correlation $\tau = \tau_{\text{PROPN}, {\text{PROPN}}}$ shows similar temperature dependence, as shown in Fig.~\ref{fig:pnas_SI_multiling_corr}. At low temperatures, the integrated correlation increases with sequence length and appears to diverge. At high temperatures, it saturates to a finite value, although the convergence is slightly slower for Spanish. We also analyze the power spectrum $S(\omega) = S_{\text{PROPN}}(\omega)$. At high temperatures, the spectrum shows a single peak at $\omega = 0$. By contrast, at low temperatures, it exhibits numerous peaks for all languages, as shown in Fig.~\ref{fig:pnas_SI_multiling_power}. From these results, we expect that the phase transition and the complex repetitive structures are observed universally across languages.

\section{Character-based sequences}
\label{app:character}

Reliable statistical analyses require mapping each text to a sequence of a moderate number of states. In the main text, we mapped texts to sequences of part-of-speech (POS) tags using the spaCy library~\cite{montani_2023_10009823}. Although we cannot fully rule out unexpected effects caused by POS tagging, we did not observe any behavior suggestive of such artifacts. To further strengthen the conclusions in the main text, we analyze sequences obtained using an alternative mapping that does not rely on POS tagging. Specifically, we map each character in a text to 1 if it is a space according to Python's isspace method, and to 0 otherwise. For the resulting binary sequences, we measured the integrated correlation and the power spectrum using the same procedures as in the main text.

\begin{figure*}[h]
    \centering
    \includegraphics[width=.7\textwidth]{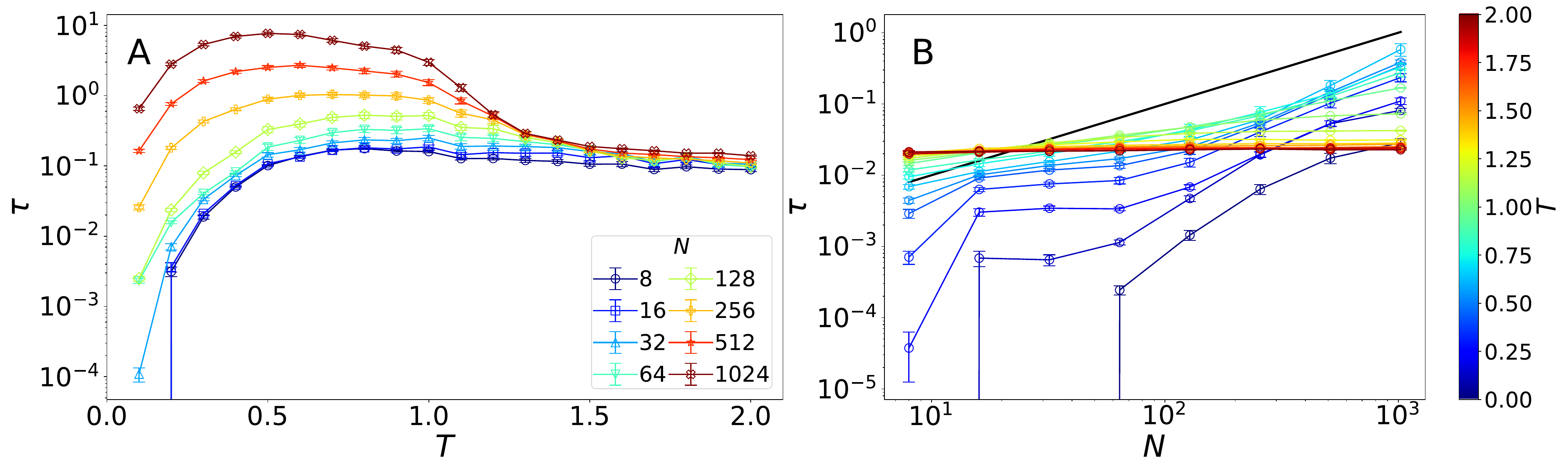}
    \caption{(A) Integrated correlation $\tau = \tau_{1, 1}$ of character-based binary sequences as a function of temperature $T$ for various sequence lengths $N$. (B) Same quantity as a function of sequence length $N$ for various temperatures $T$. The black line represents a line proportional to $N$. These results suggest that a phase transition occurs near $T \approx 1$, consistent with the findings for POS sequences.}
    \label{fig:SI_character_corr}
\end{figure*}

\begin{figure*}[h]
    \centering
    \includegraphics[width=.7\textwidth]{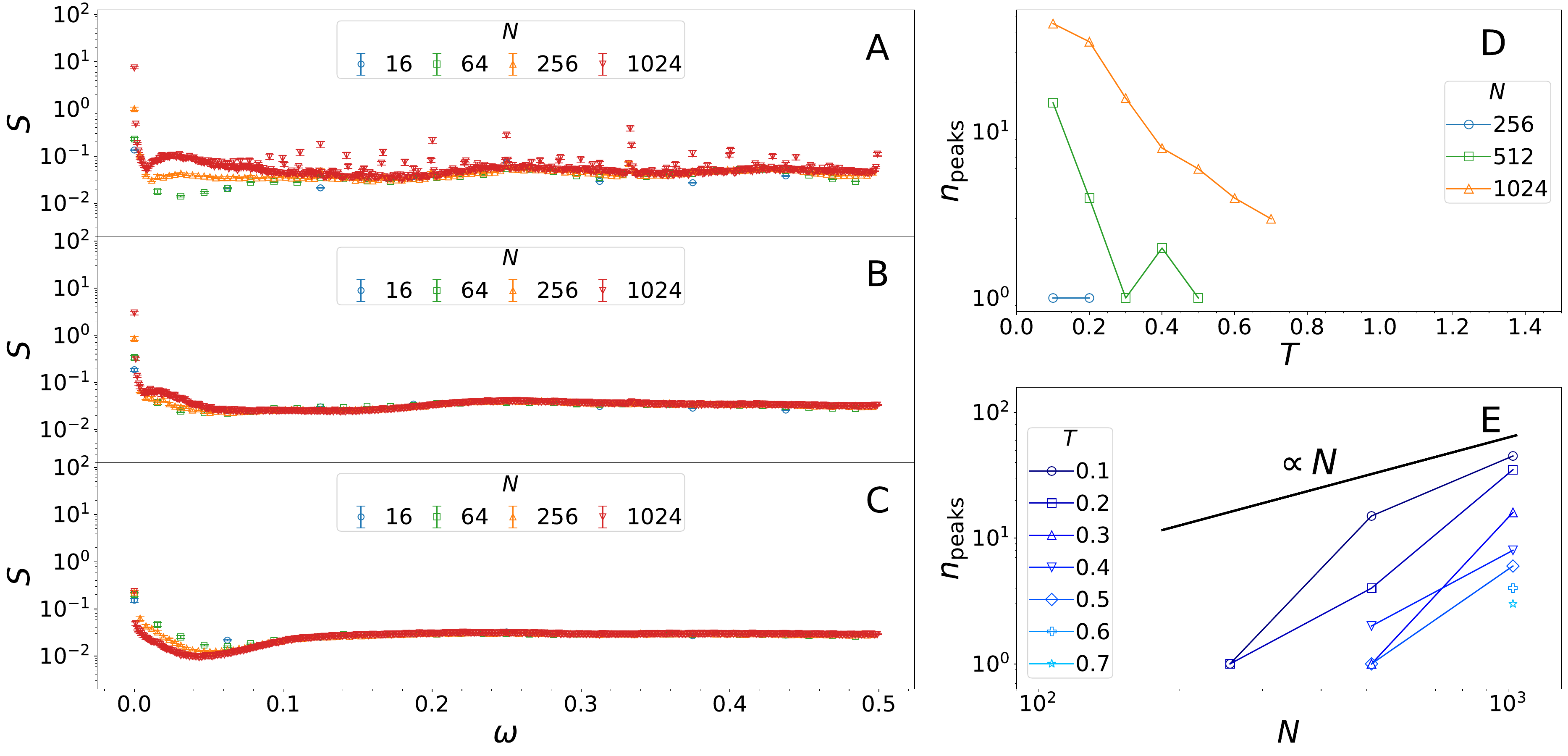}
    \caption{(A--C) Power spectrum $S = S_{1}$ of character-based binary sequences as a function of $\omega$ at (A) $T=0.6$, (B) $T=1$, and (C) $T=1.4$. (D) Number of peaks in the power spectrum as a function of $T$ for various sequence lengths $N$. (E) Same quantity as a function of sequence length $N$ for various temperatures $T$. The black line represents a line proportional to $N$. The number of peaks at low temperatures increases with sequence length, suggesting that the low-temperature regime exhibit complex repetitive structure, as in POS sequences.}
    \label{fig:SI_character_power}
\end{figure*}

The integrated correlation $\tau_{1,1}$ is shown in Fig.~\ref{fig:SI_character_corr}, where the sequence length $N$ is defined as the number of characters in a sequence. The integrated correlation continues to increase with $N$ at low temperatures but saturates at high temperatures. This behavior is consistent with that observed in the main text. Figure~\ref{fig:SI_character_power} shows the power spectrum $S_1$. At low temperatures, numerous peaks appear, and the number of peaks, $n_{\text{peaks}}$, appears to diverge. Again, the results are consistent with those in the main text. These results demonstrate that the findings in the main text are not artifacts of POS tagging.

\section{Integrated correlations for other POS pairs}
\label{app:integrated}

\begin{figure*}[h]
    \centering
    \includegraphics[width=\textwidth]{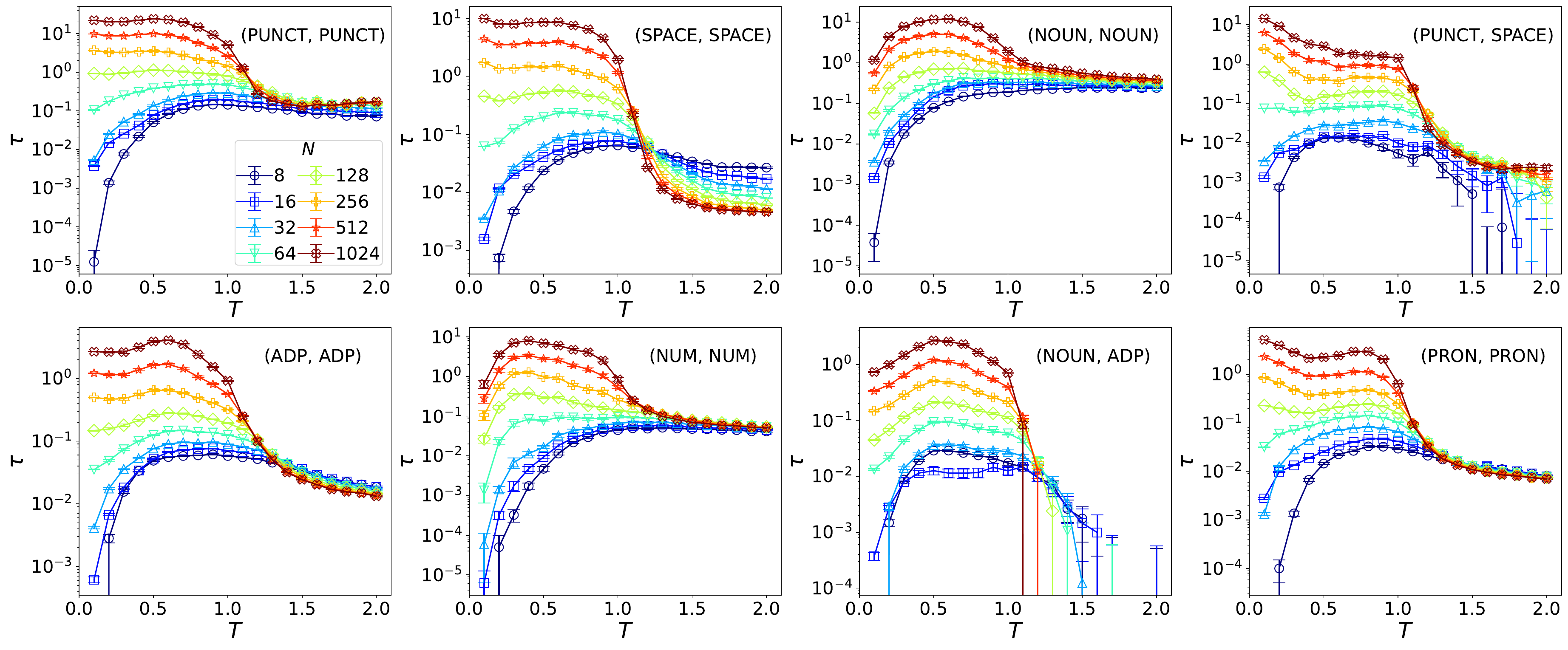}
    \caption{Integrated correlation $\tau_{ab}$ for the eight POS pairs $(a,b)$ with the largest contributions. For all pairs, the integrated correlation continues to increase at low temperatures and saturates at high temperatures, indicating that the phase transition occurs.}
    \label{fig:SI_corr_POSpairs}
\end{figure*}

The integrated correlation $\tau_{ab}$ depends on the POS pair $(a,b)$. In the main text, we observed the phase transition by analyzing this quantity for the pair $(a,b) = (\text{PROPN}, \text{PROPN})$, which gives the largest contribution. Here, we show that the phase transition is also observed for other POS pairs with large contributions. Figure~\ref{fig:SI_corr_POSpairs} shows the integrated correlations for the eight pairs with the largest absolute values of $\tau_{ab}$ at $T = 1$ and $N = 1024$, excluding $(\text{PROPN}, \text{PROPN})$. Note that $(a,b)$ and $(b,a)$ are counted as the same pair because $\tau_{ab} = \tau_{ba}$. For all pairs, the integrated correlation increases with the sequence length $N$ at $T \lesssim 1$, indicating divergence as $N \to \infty$, whereas it saturates to a finite value at $T \gtrsim 1$. This behavior indicates that the phase transition occurs near $T = 1$ for these POS pairs, as in the case of $(\text{PROPN}, \text{PROPN})$.

\section{Dynamics}
\label{app:dynamics}

\begin{figure*}[h]
    \centering
    \includegraphics[width=.9\linewidth]{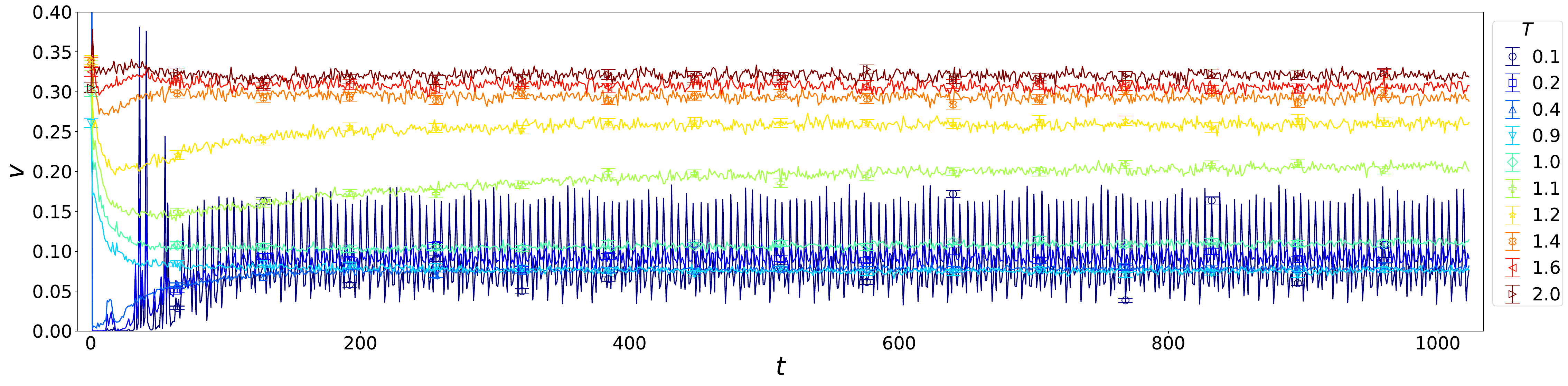}
    \caption{Probability $v(t) = v_{\text{PROPN}}(t)$ that the $t$-th tag is PROPN as a function of time $t$ at various temperatures, where the sequence length is $N=1024$. The transient time to reach the stationary state becomes extremely longer near $T_c$, suggesting the critical slowing down.}
    \label{fig:pre_dynamics}
\end{figure*}

The correlation observed in the main text suggests that critical slowing down occurs, that is, the convergence to the steady state becomes significantly slower near $T_c$. To examine this expectation, we observe how the POS tags evolve with time $t$. Specifically, we calculate the probability of POS tag at time $t$ being $a$:
\begin{equation}
    v_a(t) = \mathbb{E} [ \delta_{y_t, a} ].
\end{equation}
Figure~\ref{fig:pre_dynamics} shows $v(t) = v_\text{PROPN}(t)$ as a function of time $t$ at several temperatures. At high temperatures, $v(t)$ rapidly reaches its limiting value. At low temperatures, the dynamics also appears to approach a stationary state rapidly, but it exhibits pronounced oscillations. The power spectrum presented in the main text indicates that numerous periodic structures coexist within these oscillations. Near the critical point between the two regimes, the transient time to the stationary state becomes longer, indicating the critical slowing down.

So far, we have focused on $v(t) = v_{\mathrm{PROPN}}(t)$. However, since $a$ takes 18 different POS tags, $v_a (t)$'s form the 18-dimensional vector $\bm{v} (t)$. To reveal the underlying dynamics in the entire 18-dimensional space, we employ principal component analysis (PCA). Specifically, we concatenate $\bm{v}(0), \ldots, \bm{v}(N-1)$ across 20 temperature points, $T = 0.1, 0.2, \ldots, 2$, into an $18 \times 20N$ data matrix and apply PCA to it.

\begin{figure*}[h]
    \centering
    \includegraphics[width=.7\linewidth]{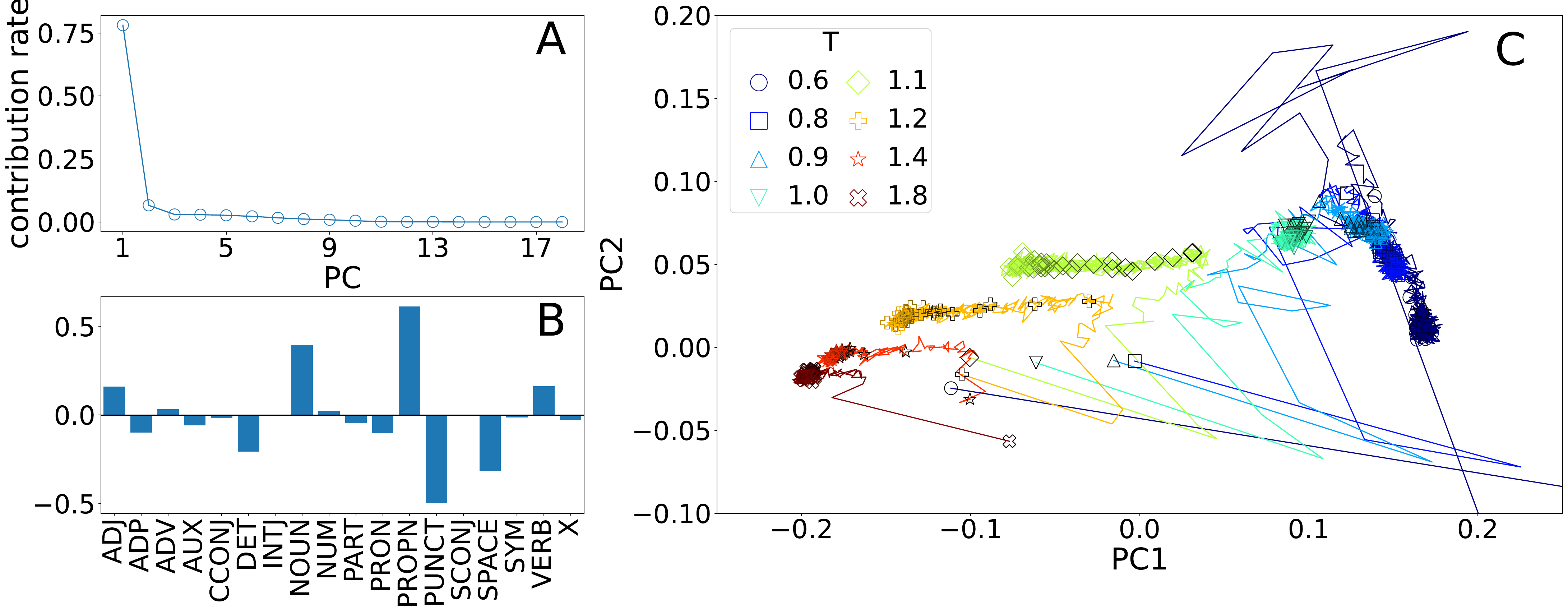}
    \caption{Results of principal component analysis (PCA) applied to the dynamics of $\bm{v}(t)$. (A) Contribution rates of the principal components. (B) Elements of PC1. (C) Dynamics of $\bm{v}(t)$ at different temperatures projected onto the two-dimensional principal component space. Markers are plotted at intervals of 32 time steps, with darker markers corresponding to smaller $t$. The dynamics slow down near the critical point.}
    \label{fig:pnas_SI_pca}
\end{figure*}

The contribution rates of the principal components, shown in Fig.~\ref{fig:pnas_SI_pca}A, indicate that the contribution of the first principal component (PC1) is sufficiently large. Figure~\ref{fig:pnas_SI_pca}B shows the elements of PC1, in which the element corresponding to PROPN has the largest absolute value, justifying our focus on the dynamics for PROPN. 

Figure~\ref{fig:pnas_SI_pca}C shows the dynamics of $\bm{v}(t)$ at different temperatures projected onto the two-dimensional principal component (PC) space. The result indicates that the stationary states reached by $\bm{v}(t)$ as $t \to \infty$ can be classified into two types. Above the critical point, the dynamics converge to one state, whereas below the critical point they converge to the other. Around the critical point, the transient time is longer than at higher or lower temperatures. These observations are consistent with the discussion of the one-dimensional dynamics of $v_{\mathrm{PROPN}}(t)$. The dynamics in the PC space further suggest that critical slowing down occurs when the two stable states potentially coexist.

\section{Power spectra for individual sequences}
\label{app:power}

In the main text, we characterized the repetitive structures of texts generated at low temperatures by analyzing the power spectrum
\begin{equation}
    S_a (\omega)
    = N \left(
        \mathbb{E} \left[
            \left|
                f_a( \omega )
            \right|^2
        \right]
        - \left|
            \mathbb{E} \left[
                f_a( \omega )
            \right]
        \right|^2
    \right).
\end{equation}
This quantity does not allow us to examine the structure of individual sequences, because it is defined as an average over sequences. Here, we introduce the power spectrum for individual sequences $(y_0, \cdots, y_{N-1})$, defined as
\begin{equation}
    \begin{aligned}
    &S_a (\omega ; y_0, \cdots, y_{N-1}) \\
    &= N \left(
        \left|
            f_a( \omega ; y_0, \cdots, y_{N-1} )
        \right|^2
        - \left|
            \mathbb{E} \left[
                f_a( \omega )
            \right]
        \right|^2
    \right).
    \end{aligned}
\end{equation}
This quantity characterizes the periodicity of individual sequences.

\begin{figure*}[h]
    \centering
    \includegraphics[width=.9\textwidth]{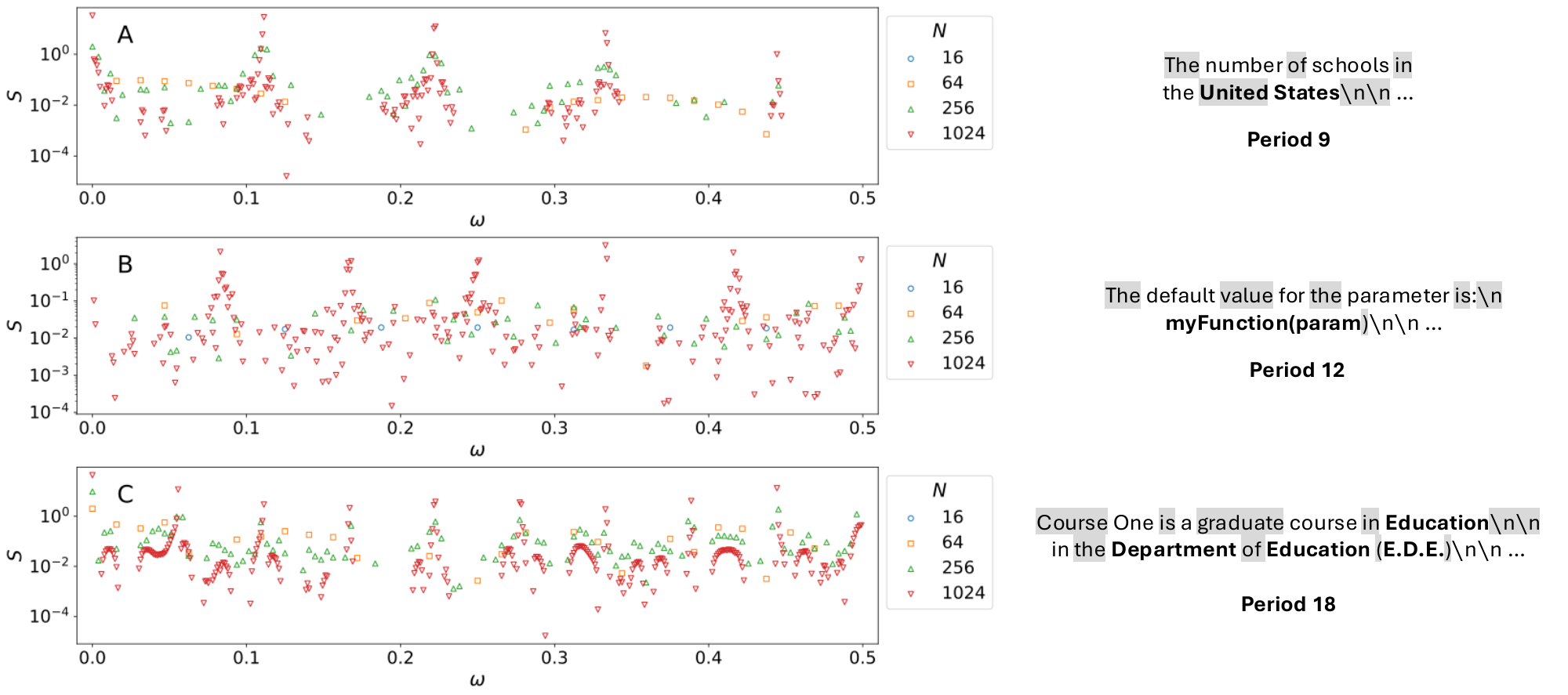}
    \caption{Power spectra for individual sequences generated by the pretrained model at $T=0.6$. The repeated pattern in each sequence is shown to the right of each panel: (A) \textit{The number of schools in the United States\textbackslash{}n\textbackslash{}n} (period 9), (B) \textit{The default value for the parameter is:\textbackslash{}nmyFunction(param)\textbackslash{}n\textbackslash{}n} (period 12), and (C) \textit{Course One is a graduate course in Education\textbackslash{}n\textbackslash{}nin the Department of Education (E.D.E.)\textbackslash{}n\textbackslash{}n} (period 18). The units segmented by the POS tagger are highlighted alternately. Units tagged as PROPN are shown in bold. The power spectrum for a sequence with period-$p$ repetitions exhibits peaks at frequencies that are multiples of $1 / p$.}
    \label{fig:SI_power_single}
\end{figure*}

In Fig.~\ref{fig:SI_power_single}, we present the power spectra for three individual sequences generated at $T = 0.6$, all of which contain repeated patterns. Figure~\ref{fig:SI_power_single}A shows the result for a sequence with repetitions of \textit{The number of schools in the United States\textbackslash{}n\textbackslash{}n}. Because the POS tagging decomposes this phrase into nine tags, this repetition gives rise to a period-9 structure. The power spectrum for this sequence exhibits clear peaks at frequencies that are multiples of $1/9$. Similarly, Figs.~\ref{fig:SI_power_single}B and C show the results for sequences with repetitions of period 12 and 18, respectively. These spectra also exhibit peaks at the corresponding frequencies, namely multiples of $1/12$ and $1/18$.

These results demonstrate that individual sequences exhibit periodic structures with different periods, depending on the repeated patterns present in the sequences. Averaging over such sequences yields the complex repetitive structure, characterized by the divergent number of peaks in the average power spectrum.

\section{Emergence of critical slowing down in training}
\label{app:slowing}

The integrated correlation for partially trained models suggests that the phase transition emerges around training step $k_c \approx 10^2$. The critical slowing down is expected to occur at the same point. To examine this expectation, we estimate the probability $v(t)$ that the $t$-th tag $y_t$ is PROPN for different training steps.

\begin{figure*}[h]
    \centering
    \includegraphics[width=.5\linewidth]{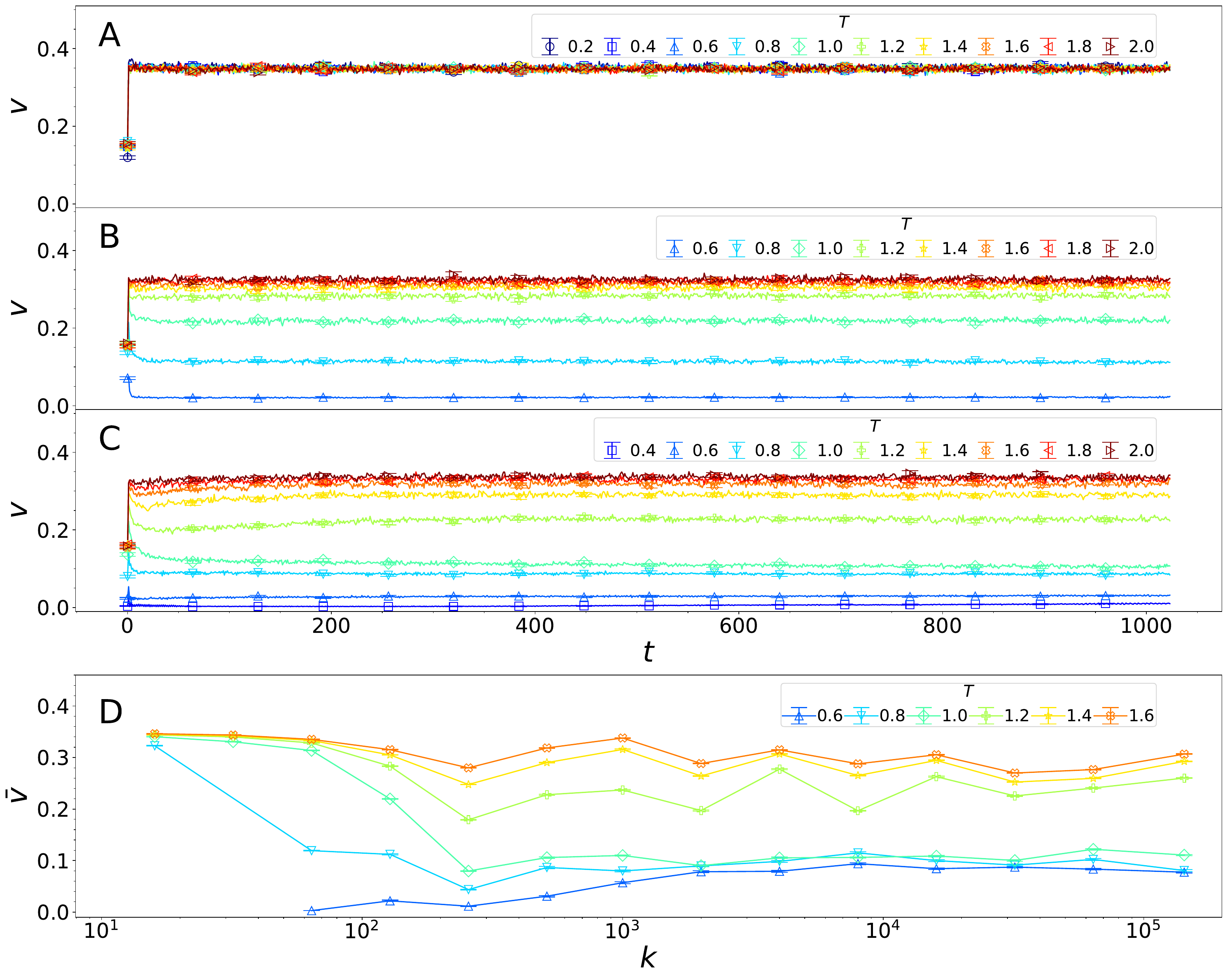}
    \caption{(A--C) Probability $v(t) = v_{\text{PROPN}}(t)$ that the $t$-th tag is PROPN as a function of time $t$ at various temperatures for training steps (A) 0, (B) 128, and (C) 512, where the sequence length is $N=1024$. (D) Estimated limiting value $\bar{v}$ as a function of training step, where $N = 1024$ and $M = 128$. The slowing down of dynamics emerges at training steps $k \approx 10^2$, consistent with the onset of the phase transition at $k_c \approx 10^2$. Several values of $v(t)$ and $\bar{v}$ for early steps at low temperatures are not shown because text generation terminates too early to obtain a sufficient number of long sequences.}
    \label{fig:par_dyn}
\end{figure*}

Figures~\ref{fig:par_dyn}A--C show the results for steps $k = 0$, 128, and 512. For the initialized model ($k = 0$), the dynamics of $v(t)$ are trivial: They rapidly converge to the same stationary state across temperatures. As the training step approaches $k_c$, the dynamics begin to depend on the temperature. At step $k = 128$, the limiting value of $v(t)$ is larger at higher temperatures and smaller at lower temperatures. With further training, the limiting values at low and high temperatures become more clearly separated, as shown for $k = 512$. In addition, the transient time required to reach the stationary state increases at intermediate temperatures, indicating the onset of critical slowing down.

To systematically examine when the critical slowing down arises, we observe how the limiting value of $v(t)$ depends on the training step $k$. We estimate the limiting value by averaging $v(t)$ over the last $M$ time points, i.e., $\bar{v} = \sum_{t = N - M}^{N - 1} v(t)/M$. Figure~\ref{fig:par_dyn}D shows this estimated limiting value as a function of $k$ at different temperatures, where we use $N = 1024$ and $M = 128$. The results show that the limiting values for low and high temperatures begin to separate around $k_c$, consistent with the expectation that the phase transition and the critical slowing down arise at the same stage of training.

\section{Statistical properties of natural language}
\label{app:natural}

We showed that texts generated by the pretrained model exhibit critical behavior. Near the critical point $T_c$, the statistical properties of these texts are expected to be consistent with those of natural languages, since power-law behavior has been observed across natural languages. To test this expectation, we measure the same statistical quantities on $1.6 \times 10^5$ texts sampled from the Pile \cite{gao2020pile}, which is used to train the Pythia models.

\begin{figure*}[h]
    \centering
    \includegraphics[width=.55\linewidth]{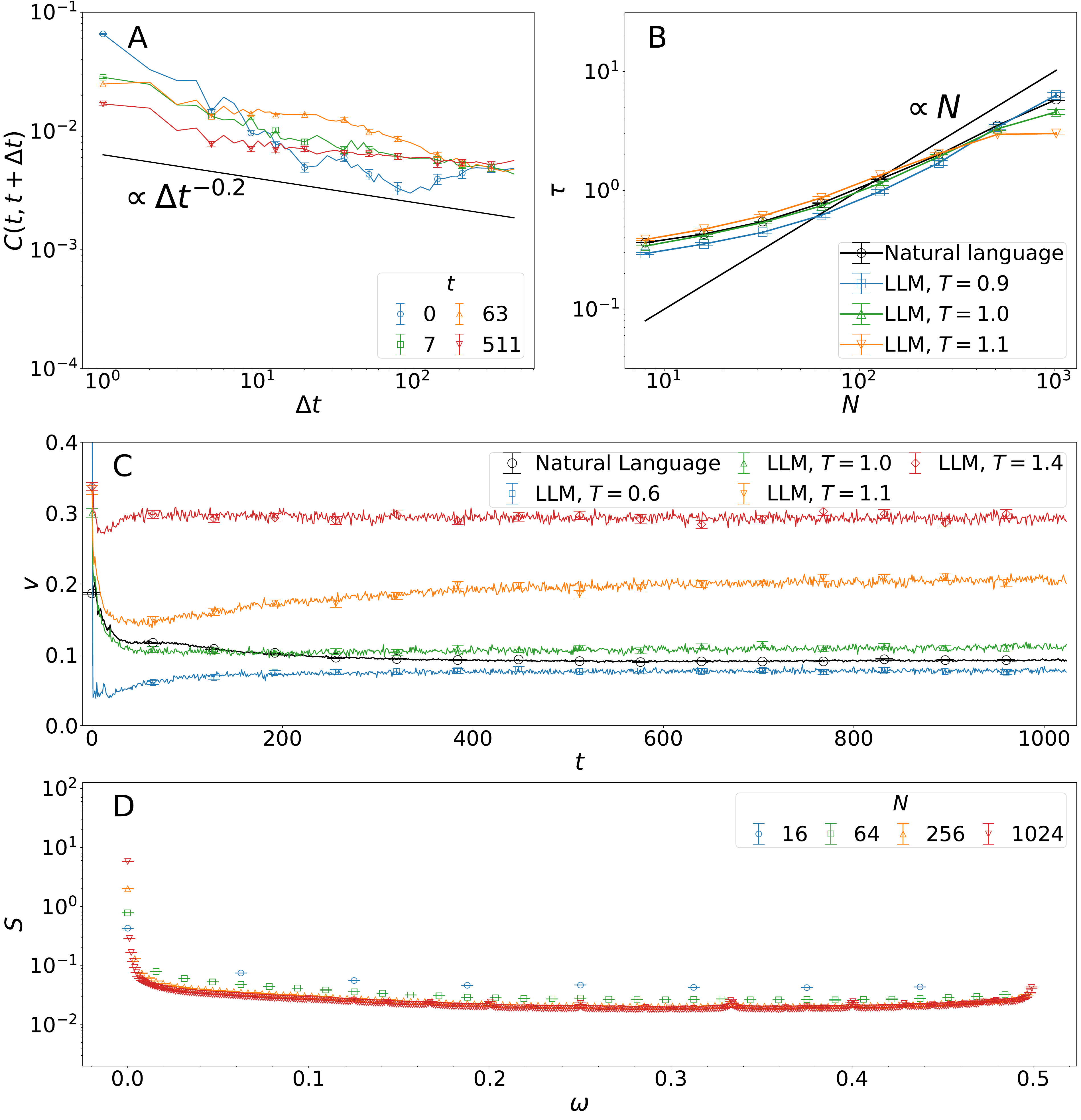}
    \caption{(A) Correlation $C(t, t+\Delta t) = C_{\text{PROPN},\text{PROPN}}(t, t+\Delta t)$ for the Pile dataset, as a function of time interval $\Delta t$, where the sequence length is $N = 1024$. The black line represents a line proportional to $\Delta t^{-0.2}$. (B) Integrated correlation $\tau = \tau_{\mathrm{PROPN},\mathrm{PROPN}}$ for the Pile, and for sequences generated by Pythia 160M at $T = 0.9$, $1$, and $1.1$, shown as a function of sequence length $N$. (C) Dynamics of $v(t) = v_{\text{PROPN}}(t)$ for the Pile, together with those for texts generated by the pretrained model at several temperatures. The sequence length is $N = 1024$. (D) Power spectrum $S = S_{\mathrm{PROPN}}$ for the Pile. For all these quantities, the Pile exhibits behaviors similar to those for the critical pretrained Pythia model.}
    \label{fig:pnas_SI_natural}
\end{figure*}

Figures~\ref{fig:pnas_SI_natural}A--D show the correlation, the integrated correlation, the dynamics, and the power spectrum for the Pile, respectively. All these quantities exhibit behavior similar to that observed for the critical pretrained model: The correlation exhibits a critical decay; the integrated correlation continues to grow at a rate similar to that of the critical model; the convergence of $v(t)$ to its limiting value is slower than in the pretrained model at both low and high temperatures; and the power spectrum exhibits a single divergent peak at $\omega = 0$ and small finite peaks at other frequencies, similar to that observed for texts generated at $T = 1$. These results confirm that the natural language dataset shares statistical properties with the critical LLM.

\clearpage
\onecolumngrid

\clearpage
\twocolumngrid

\bibliography{transition_in_LLM_PNAS_arXiv_20260606}

\end{document}